\pdfoutput=1
\documentclass[sigconf, nonacm]{acmart}
\usepackage{pvldb}
\setkeys{acmart.cls}{balance=false}
\usepackage{booktabs}
\usepackage{tabularx}
\usepackage{array}
\usepackage{amsmath}
\usepackage{placeins}
\usepackage{xltabular}
\usepackage{pifont}

\usepackage{algorithm}
\usepackage{algpseudocode}
\hypersetup{hidelinks}
\AtBeginDocument{\color{black}}
\newcommand{\system}{SpecBridge}
\newcommand{\extref}[1]{\ref{#1}}

\newcommand{\sql}[1]{\textnormal{\texttt{#1}}}
\newcommand{\req}[1]{{\color{black!65}\ding{\numexpr201+#1\relax}}}
\newcommand{\method}[1]{\textsc{#1}}
\newcolumntype{Y}{>{\raggedright\arraybackslash}X}
\graphicspath{{figures/}}
\title{Text-to-SQL Models Make Choices Without Saying So}
\author{Yichen Xu}
\affiliation{%
  \institution{University of California, Berkeley}
  \city{Berkeley}
  \state{CA}
  \country{USA}
}
\email{yichen\_xu@berkeley.edu}
\author{Shipei Lin}
\affiliation{%
  \department{School of Data Science}
  \institution{The Chinese University of Hong Kong, Shenzhen}
  \city{Shenzhen}
  \country{China}
}
\email{225045017@link.cuhk.edu.cn}
\author{Chenhao Ma}
\affiliation{%
  \department{School of Data Science}
  \institution{The Chinese University of Hong Kong, Shenzhen}
  \city{Shenzhen}
  \country{China}
}
\email{machenhao@cuhk.edu.cn}
\begin{document}
\settopmatter{printacmref=false}
\begin{abstract}
A text-to-SQL answer can run, return a plausible number and still compute something other than what the analyst meant. This can happen when the request leaves a calculation choice open, such as whether to count rows or distinct entities, and the model settles it without saying so. Standard benchmarks cannot detect this, because they score each question against a single reference interpretation. We introduce \system{}, which tests such a choice by varying it alone: it poses three otherwise identical requests, one leaving the choice open and one per interpretation, and scores each answer by running it. The databases are real, and the choices are ones that made models fail on existing benchmarks. On 204 hand-validated tests, four language models usually compute each interpretation when asked for it, yet for the open request return both in only 15--58\% of the tests where they can, often without mentioning the choice. The gap lies in reporting, not in computing. Benchmarks even penalize the alternative: on BIRD questions where such a choice changes the answer, over two in five of the answers marked wrong are right under the other interpretation. Because the gap lies in reporting, a system can narrow it without a stronger model: it first decides whether a request leaves a choice open, then answers each interpretation in its own call. On our tests and on AMBROSIA, an existing benchmark of ambiguous questions, this approach raises the share of open requests that get every interpretation by 29--63 points over the single answer that production assistants give, without lowering precision on clear requests. It also needs fewer user turns than asking a clarifying question.
\end{abstract}
\keywords{text-to-SQL, database interfaces, evaluation, controlled tests, ambiguity, language models}
\maketitle
\noindent\textbf{Artifact availability:} the code, data and other artifacts are available at \url{https://github.com/Xyc-arch/specbridge-artifact}.\par
\section{Introduction}
\label{sec:intro}
Analysts increasingly ask for numbers in natural language and get them from a text-to-SQL system, often without reading the SQL. For these users a query that fails is safe, because the error is visible. The dangerous case is a query that runs and returns a plausible number, computed differently than the analyst meant. Small choices in how a number is computed can change a published finding. An influential study reported that high public debt slows growth. It first averaged growth within each country, then across countries, so a country with one high-debt year counted as much as a country with many. With corrected data and each year counted equally, high-debt countries grew 2.2\% a year, not $-0.1\%$~\cite{herndon2014debt}. Table~\extref{tab:cases} lists more cases across six domains. A text-to-SQL system makes such choices on many requests, and the result alone does not show them. We call such a choice, settled without telling the user, a silent choice. The answer is correct for one interpretation, so checks that look for wrong SQL~\cite{chen2023error,alrashed2026verification} do not catch it.

\begin{figure}[t]
\centering
\includegraphics{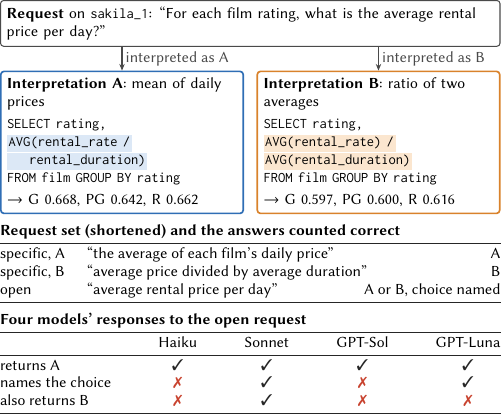}
\caption{One request, its two correct interpretations (shaded: the SQL that differs), its request set, and four models' responses.}
\Description{A request on the film table leads to two interpretations, A (mean of daily prices) and B (ratio of means): the SQL expression
that differs, and their different per-rating values. Below, the three requests of the request set, and a table of four models' recorded responses to the open request.}
\label{fig:teaser}
\end{figure}

Take ``the average rental price per day'' for each film rating (Figure~\ref{fig:teaser}). It can be computed in two ways, which we call interpretations. Interpretation A averages each film's daily price, and interpretation B divides the average price by the average rental duration. For two films priced at \$6 for two days and \$6 for six days, A gives $(\$3 + \$1)/2 = \$2.00$ per day and B gives $\$6/4 = \$1.50$. Both are simple SQL, and on Spider's \texttt{sakila\_1} database the two interpretations differ for every rating. We test four models: Claude Haiku~4.5, Claude Sonnet~5.5, GPT-6 Sol and GPT-6 Luna (Haiku, Sonnet, GPT-Sol and GPT-Luna below). For the request as worded, which leaves this choice open, all four return A. Sonnet and GPT-Luna say that there is a choice, but only Sonnet also gives B, the ratio of two averages.

A reliable text-to-SQL system must do three things: \req{1}~compute the interpretation that a specific request asks for, \req{2}~report a choice that a request leaves open together with the other interpretation, and \req{3}~stay quiet about a choice that the wording already settles. Production assistants handle open choices by stopping. Snowflake's Cortex Analyst rejects a question its classifier marks ambiguous~\cite{sehrawat2024cortex}, Databricks' Genie asks a clarifying question where its authors' instructions say so~\cite{genie2025}, and Microsoft's Fabric data agent leaves it to instructions its authors write~\cite{fabric2026agent}. When these assistants do answer, they return one interpretation and state it~\cite{cortexrest2025}. By default, none of them meets \req{2}.

Benchmarks measure \req{1}--\req{3} only in part. Spider and BIRD measure only \req{1}. They score each question against one reference, even when the question leaves a choice open. A model that guesses the authors' interpretation thus scores as well as one that reports the choice. Ambiguity benchmarks~\cite{ambiqt2023,ambrosia2024} add \req{2}, and AMBROSIA also adds \req{3}. Three gaps remain (Table~\ref{tab:comparison}). First, both create their choices: AmbiQT edits Spider's schemas, and AMBROSIA generates synthetic databases. These choices lie in the schema or the wording, such as which column a word means. Choices in how a number is computed, as in the debt study, are not among them. Yet such choices arise naturally in BIRD and Spider questions, and the references settle them by convention. Second, both score only the SQL a model returns. A user who does not read SQL learns of a choice only if the model says so. Third, their scores do not show why a model misses an interpretation. The model may be unable to compute it, or it may compute it but not report it. The first calls for a stronger model; the second does not.

\begin{table}[t]
\caption{\system{} and the two closest benchmarks of ambiguous text-to-SQL. Schema: synonym or split columns and tables; wording: scope, attachment, vagueness; calculation: ties, counting units, missing values, aggregation level. Possible: AMBROSIA's data allow it, but its scores do not (Section~\ref{sec:disclosure}).}
\label{tab:comparison}
\centering\footnotesize
\providecommand{\cm}{\ensuremath{\checkmark}}\providecommand{\xm}{\ensuremath{\times}}
\setlength{\tabcolsep}{3.5pt}
\begin{tabular}{@{}lccc@{}}
\toprule
& AmbiQT~\cite{ambiqt2023} & AMBROSIA~\cite{ambrosia2024} & \system{}\\
\midrule
Where the choices lie & schema & wording & calculation\\
\midrule
Benchmark databases, unmodified & \xm & \xm & \cm\\
Scores the explanation & \xm & \xm & \cm\\
Separates computing from reporting & \xm & possible & \cm\\
Penalizes raising a settled choice & \xm & \cm & \cm\\
\bottomrule
\end{tabular}
\end{table}
Our key idea is a controlled test: three requests that differ in the choice and in nothing else. Two are specific, one asking for interpretation A and one for B. The third, the open request, leaves the choice open. With the two reference queries, the three requests form a request set (Figure~\ref{fig:teaser}). All three use the same prompt. It asks the model to state any assumption that the request does not settle and to give SQL for other interpretations. We score the SQL by execution and the explanation by a judge model. Because only the choice differs, the test separates three kinds of failure. A wrong answer to a specific request is a failure to compute (\req{1}). On the open request, missing an interpretation that the model computes when asked for it alone is a failure to report (\req{2}). Raising the choice on a specific request is a false alarm (\req{3}), because its wording settles the choice.

Building request sets at scale is hard for three reasons, and our pipeline, \system{}, addresses each in turn. First, a specific request must state every requirement of its query, so that a wrong answer means a failure to compute. BIRD and Spider questions often leave a requirement unstated, so they cannot serve as specific requests. \system{} therefore generates new SQL queries on unmodified benchmark databases, and code describes each query exactly in words (Section~\ref{sec:construction}). Second, a test reflects real use better when its choice is one that actual questions leave open, but no benchmark lists such choices. \system{} finds them in benchmark failures. When a model fails a BIRD or Spider question but answers correctly from an exact description of its reference SQL instead, the wording is to blame, sometimes because it leaves a choice open (Section~\ref{sec:library}). Third, the three requests must differ only in the choice, but when a model writes them, it often changes something else too, such as a filter. \system{} therefore has the model write all three from one plan that lists what they share, and checks every set with code and two reviewer models (Section~\ref{sec:apply}). Two authors then reviewed the 307 sets that pass, without seeing model results, and kept 232. Results use the 204 of them on choice types where BIRD and Spider references take both interpretations (Section~\ref{sec:specresults}). Execution, not a model, decides every accuracy score in this paper.

\textbf{Finding 1: Models make choices without saying so} (Section~\ref{sec:disclosure}). On 204 accepted request sets (Figure~\ref{fig:disclosure}), the models mostly compute what is asked (\req{1}) and almost never raise a settled choice (\req{3}), but leave open choices unreported (\req{2}). For the open request, they return both interpretations in only 15--58\% of the sets they can compute, and they name the choice in only 38--70\% of all sets. Saparina and Lapata~\cite{saparina2025disambiguate} observed that models usually give one preferred interpretation of an ambiguous question. Request sets show that they can usually compute the others: in 72--94\% of the sets where a model misses an interpretation, the same model computes it correctly when asked for it alone. Most misses are thus failures to report, not failures to compute.

\textbf{Finding 2: Benchmarks cannot see unreported choices} (Section~\ref{sec:diagnosis}). When a BIRD or Spider question leaves a choice open, its reference SQL still takes one interpretation, by the convention of the group that wrote it. BIRD's and Spider's own authors return one row when several rows tie, but questions that Spider imported from two older datasets usually return all of them. On BIRD, 156 development questions, one in ten of all, change their answer under the other interpretation. On 79 such questions where both interpretations are valid, accepting both raises the four models' accuracy from 25\% to 57\%. For these questions, a benchmark score measures agreement with the benchmark's authors, not correctness, and cannot show whether a model reports the choice.

\textbf{Finding 3: Deciding first makes reporting more reliable} (Section~\ref{sec:method}). Because most misses are failures to report, not failures to compute, a method can return many of the missed interpretations without a stronger model. Phase 1 of our method, \method{Decide-then-answer}, first decides whether a request leaves a choice open and then answers each interpretation of an open request in its own call. Compared with the single answer that production assistants give, it returns every interpretation for 29 to 63 points more of the open requests on AMBROSIA and our request sets, without lowering precision on clear requests. In a simulated conversation, it also needs 35.0 to 40.8 fewer user turns per 100 requests than asking a clarifying question. Once the user picks an interpretation, today's systems and methods, such as LangChain~\cite{langchain2024crc} and \method{DtP}~\cite{saparina2025disambiguate}, rewrite the request, which usually keeps the choice but can change a formula or drop an output column. Phase 2 instead keeps the request and adds the pick as one clause, which computes 8.5 to 10.5 points more interpretations correctly.

In summary, we make three contributions: (1) request sets, controlled tests that change one calculation choice and so separate computing from reporting, with a generator that builds them on real databases and a library of choices mined from benchmark failures (Sections~\ref{sec:framework}--\ref{sec:evaluation}); (2) findings that models compute more than they report and that benchmarks cannot see this, on four models across our request sets, AMBROSIA and BIRD (Sections~\ref{sec:disclosure}--\ref{sec:diagnosis}); and (3) a two-phase method that reports open choices and carries the user's pick, tested against prompting, sampling, \method{DtP} and production behavior rebuilt from documentation (Section~\ref{sec:method}). We release the generator, the 307 request sets with the authors' verdicts, all model responses, and a package that reproduces every number in this paper without new model calls.
\section{\system{} Overview}
\label{sec:framework}
\begin{figure}[t]
\centering
\includegraphics{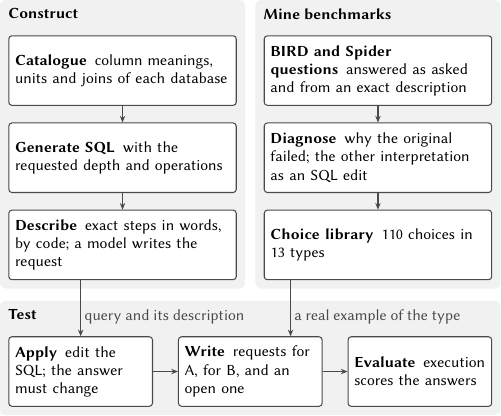}
\caption{The \system{} pipeline.}
\Description{Two lanes of three boxes, Construct (catalogue, generate SQL, describe) and Mine (benchmark questions, diagnose,
decision library), both feed a bottom lane: apply a choice, write three requests, evaluate.}
\label{fig:pipeline}
\end{figure}

\system{} is named for its central step, which bridges an exact specification of each query, written by code, and the natural requests a user would write. It turns a database into request sets in two lanes that meet in one step (Figure~\ref{fig:pipeline}). The first lane generates queries whose computation is known exactly. The second lane mines, from benchmark failures, the choices that real requests leave open. The last step applies one such choice to a generated query and writes the three requests. We follow two request sets through this process (Section~\ref{sec:worked}) and then describe each step.
\subsection{Two examples}
\label{sec:worked}
\label{sec:test}
\begin{figure}[t]
\centering
\includegraphics{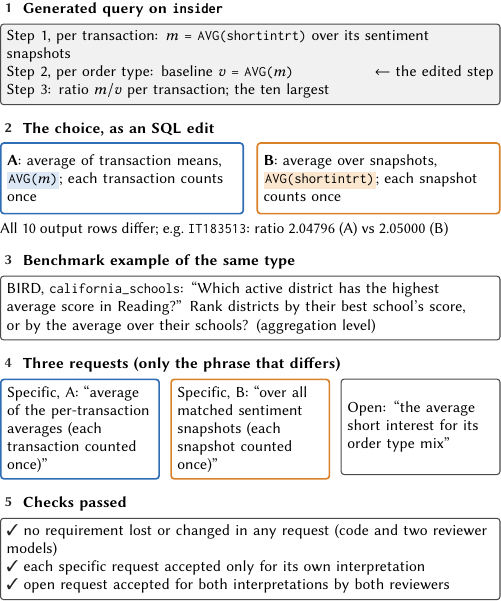}
\caption{A request set on \texttt{insider}, from the generated query to its checks. Shaded: the SQL that differs.}
\Description{Five stacked boxes: the generated query with its edited step, the two interpretations and their executed difference, the
retrieved library example, the three requests, and the checks passed.}
\label{fig:worked}
\end{figure}

A request set fixes a database and every requirement of a query but one choice. The film request of Figure~\ref{fig:teaser} asks for each rating's average replacement cost and daily rental price, and compares that daily price with the average across ratings. Only the daily price is open:
\begin{equation}
 A=\operatorname{AVG}(\text{price}/\text{days}),\qquad
 B=\frac{\operatorname{AVG}(\text{price})}{\operatorname{AVG}(\text{days})}.
\end{equation}
A averages each film's daily price; B divides two averages and so gives more weight to longer rentals. The specific request for A asks for the average of per-film daily prices, the one for B asks for the average price divided by the average duration, and the open request asks for the ``average rental price per day.'' The rest of the wording is shared, and code checks that the two queries return different results on the database; a choice that does not change the answer is not tested there. \label{sec:additionalexample}The second example (Figure~\ref{fig:worked}) is a three-step query on the \texttt{insider} database. It averages short-interest measurements for each transaction, compares each transaction with a baseline for its order type, and returns the ten largest ratios. The choice lies in the baseline. Interpretation A averages the transaction means, so each transaction counts once, while B averages the underlying snapshots, so a transaction with more snapshots weighs more. The open request asks for ``the average short interest for its order type mix,'' and all ten output rows change between the interpretations.

\subsection{Generating queries}
\label{sec:construction}
\label{sec:querygen}
\label{sec:writing}
Generation starts from the database, not from a benchmark, so it works on databases nobody has annotated. One model review of the schema and data first builds a catalogue of column meanings, units and table relationships; after that, generation calls no model. A query starts from counts, sums or averages per entity, per group or overall. Later steps can compare a value with a broader average, take a share of a total, combine compatible quantities, aggregate again, or keep the top entities per group. We control two properties of each query, its arithmetic depth (the nesting of arithmetic expressions) and its step depth (the number of dependent query stages). Rules on column types decide which quantities can be combined or summed, and data checks reject empty answers, constant or redundant quantities, non-positive denominators and too-small groups. The generator looks ahead and drops a partial query as soon as it can no longer reach the requested depths, so no attempt is spent on a dead end, and every delivered query has exactly the arithmetic and step depths it was asked for.

Code then describes each query exactly in words. The text covers source rows, joins, unmatched rows, filters, groups, formulas, missing values, ordering, limits, ties and SQLite conventions, and the same SQL always gives the same text; we call it the exact description. Because the exact description states every requirement, a model that fails from it cannot compute the query, while a model that succeeds from it but fails from a natural request was misled by the wording. Section~\ref{sec:library} builds on this contrast. Sonnet turns the exact description into one natural request using the column meanings. Claude Opus~5.5 (Opus below) checks the request for missing, changed or added requirements, and code checks how stored values are written. A failing request gets one revision with the reviewer's notes and is dropped if it fails again.

\begin{figure*}[t]
\centering\includegraphics{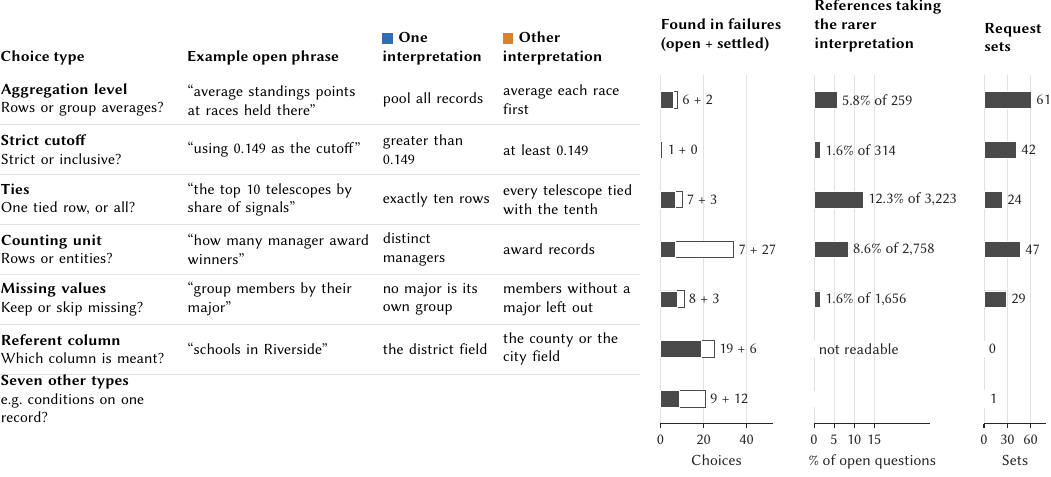}
\caption{Choice types: an example open phrase with its two interpretations, choices found in benchmark failures (gray: open; white: settled), open BIRD and Spider questions whose reference takes the rarer interpretation, and accepted request sets.}
\Description{One row per choice type with an example phrase, its two interpretations in blue and orange, and three bar panels: choices found in failures, references taking the rarer interpretation, and request sets.}
\label{fig:choices}
\end{figure*}

\subsection{Mining choices from benchmark failures}
\label{sec:library}
The choices come from real benchmark failures. For each BIRD and Spider question, we compare a model's answer to the original question (with BIRD's evidence) with its answer to the exact description of the reference SQL. When only the description is answered correctly, the wording caused the failure, not the model's SQL, and we classify the reason as output columns, a different column or value, an unstated requirement, numerical precision, or an open choice. For an open choice, the other interpretation is written as an edit of the reference SQL, and both queries are executed. Opus and Sonnet judge each interpretation independently, and disagreements are kept. The library keeps choices of meaning, both open ones, where the wording fits both interpretations, and settled ones, where it selects one. For ``What is the total number of molecules with double bonded oxygen?'', both reviewers accept only the interpretation in which the oxygen atom is part of the double bond, although the reference SQL counts every molecule that contains an oxygen atom and a double bond. Figure~\ref{fig:choices} groups the library by choice type. The largest group of open choices concerns which column a phrase refers to, and half of the settled ones concern the counting unit.

The same choices appear in benchmark references, and the references do not agree on them. We say that a question leaves a choice open, or is open, when neither its wording nor BIRD's evidence contains a word that settles it, such as ``all'' for ties or ``distinct'' for counts. Code reads which interpretation each reference takes, and Figure~\ref{fig:choices} counts, per choice type, the open BIRD and Spider questions and the share whose reference takes the rarer interpretation. Every case of the rarer interpretation was reviewed and only confirmed cases are counted, so these shares are conservative. For example, 12.3\% of open tie questions return every tied row, and 8.6\% of open questions that count over a join count distinct entities. Scoring one interpretation thus treats a convention as the only answer.

\subsection{Building request sets}
\label{sec:apply}
\label{sec:filters}
A request set needs a choice that matters on the data, so code first reports where a generated task's columns have missing values, repeated join matches, distinct values or ties. Opus proposes one choice from the library and an exact SQL edit for it, and code checks that the edit applies, runs, changes the answer and can still be described exactly; a failed proposal gets one repair. Opus then writes the request set from a plan that lists the shared requirements and the one differing choice. It writes the request for A, edits it for B, and writes the open request. A library example of the same type may guide the wording, but never supplies the SQL. Opus and Sonnet, as two reviewers, check that the requests keep their requirements and judge each of the three requests against each interpretation. A set is kept only when both reviewers accept each specific request for its own interpretation alone and at least one accepts the open request for both interpretations. Five library types concern a filter, such as whether a cutoff is strict. Generated queries seldom filter, so for these types the generator adds one condition chosen from the data. Figure~\ref{fig:choices} gives an example of each type. The pipeline keeps 307 request sets on 12 databases, and 204 of them pass the authors' review and a check against real references (Section~\ref{sec:specresults}).

\section{Measures}
\label{sec:metrics}
We score what a model computes by execution and what it says by a model judge. We report four measures: accuracy for \req{1}, both returned and named for \req{2}, and false alarms for \req{3}. A specific request selects one interpretation, and the open request fits both. A choice is settled when the wording selects one interpretation. We execute every returned query and compare its result with the reference result, with numerical tolerance and with rows tied at a cut-off treated as interchangeable; an execution error scores zero, as in Spider and BIRD scoring. For benchmark questions we also report a score that allows extra or reordered output columns.

For the open request, the model returns a main query and up to two alternatives. The model returns both interpretations when these queries include one with the result of A and one with the result of B. We report this share over all sets and over the sets in which the same model answers both specific requests correctly, where a miss cannot be a failure to compute. A missed interpretation is computable when the same model answers the specific request for it correctly. A model names the choice when its explanation says that the choice is open, whether or not it gives SQL for the other interpretation. A judge model, which does not know which model answered, decides whether a response names the choice and whether its stated interpretation matches its SQL. For a specific request, offering the other interpretation, or calling the choice open, is a false alarm; an alternative on another choice, which the set does not test, is not. The measures check one another. Returning both interpretations alone would reward a model that adds every alternative it can think of, and the false alarms charge for that; naming alone would reward a warning without an answer, and returning both interpretations charges for the missing query (Table~\ref{tab:scoring}).

\begin{table}[t]
\caption{Four recorded responses and their scores.}
\label{tab:scoring}
\centering\footnotesize\setlength{\tabcolsep}{3pt}\renewcommand{\arraystretch}{1.15}
\providecommand{\cm}{\ensuremath{\checkmark}}\providecommand{\xm}{\ensuremath{\times}}
\begin{tabularx}{\columnwidth}{@{}>{\raggedright\arraybackslash}p{1.6cm}>{\raggedright\arraybackslash}p{1.45cm}Yccc@{}}
\toprule
Request, model & Main SQL & Alternative SQL & Names & Both & False alarm\\
\midrule
Film, open, GPT-Sol & A: mean of daily prices & none & \xm & \xm & \\
Film, open, GPT-Luna & A: mean of daily prices & none & \cm & \xm & \\
Awards, open, Sonnet & A: \sql{COUNT(*)} & B: \sql{COUNT(DISTINCT player\_id)} & \cm & \cm & \\
Districts, specific, Sonnet & A, as asked & none & & & \cm\\
\bottomrule
\end{tabularx}
\end{table}

\textbf{The scores are sound by construction.} The design lets execution decide every score. Code describes each query in full (Section~\ref{sec:construction}), and each specific request is written from that description, so it states every step, the choice included. A specific request therefore has one correct result, and any other result is wrong. The open request drops only the choice phrase, so it differs from the specific requests in the choice alone. Its two interpretations are not ours: BIRD and Spider references take each of them on open questions, and we keep only choice types for which at least 1\% of references take the less common one (Section~\ref{sec:specresults}). We therefore measure precision, the share of returned queries with the correct result, on specific requests only. On the open request we ask only whether both A and B are returned, and other answers never count against the model. If a choice admits a third interpretation, we do not score it; this can only lower the true share of open requests answered with every interpretation, so the gaps we report are conservative.

On AMBROSIA, each ambiguous question comes with two or more interpretations (three for 30 of the 90 we test), and an interpretation is found when some returned query gives its result. Recall is the share of interpretations found, and AllFound is the share of questions with every interpretation found~\cite{ambrosia2024}. Precision on an unambiguous question is the share of returned queries with its one correct result, so extras lower it.
\section{Experimental Setup and Test Checks}
\label{sec:evaluation}
This section gives the setup and checks that the tests change only the choice.

\subsection{Experiment settings}
\label{sec:setup}
\textbf{Data.} Request sets use 12 databases, four each from Spider, BIRD development and LiveSQLBench\footnote{\url{https://github.com/bird-bench/livesqlbench}}. The generated tasks request arithmetic depth 1--2 and step depth 3--6, and each database contributes 12 to 24 accepted request sets. Table~\ref{tab:data} lists the data.

\begin{table}[t]
\caption{The data and where each is used.}
\label{tab:data}
\centering\footnotesize\setlength{\tabcolsep}{3pt}
\begin{tabularx}{\columnwidth}{@{}>{\raggedright\arraybackslash}p{1.95cm}>{\raggedright\arraybackslash}X>{\raggedright\arraybackslash}p{1.05cm}@{}}
\toprule
Use & Data & Sections\\
\midrule
Request sets & 12 databases, four each from Spider (baseball, Sakila, customers-and-invoices, Chinook), BIRD (financial, student club, thrombosis prediction, Formula 1) and LiveSQLBench (archeology, insider, crypto, alien) & \ref{sec:apply}, \ref{sec:disclosure}, \ref{sec:method}\\
Mining choices & 481 BIRD development questions on five databases and 150 Spider development questions & \ref{sec:library}\\
Held-out test & 100 BIRD and 50 Spider development questions, never used for mining & \ref{sec:diagnosis}\\
Conventions & every open BIRD and Spider question, training and development & \ref{sec:library}, \ref{sec:defaults}\\
Answer changes & every open BIRD development question & \ref{sec:impact}\\
Method & AMBROSIA test split, ambiguous and unambiguous questions & \ref{sec:disclosure}, \ref{sec:method}\\
\bottomrule
\end{tabularx}
\end{table}

{\sloppy\textbf{Models.} We evaluate Haiku, Sonnet, GPT-Sol and GPT-Luna (\texttt{claude-haiku-4-5-20251001}, \texttt{claude-sonnet-5-5}, \texttt{gpt-6-sol}, \texttt{gpt-6-luna}). Opus writes the request sets, reviews them with Sonnet (Section~\ref{sec:apply}), reviews the natural requests that Sonnet writes for generated tasks (Section~\ref{sec:construction}), and judges the explanations; Opus is not evaluated. Models see the schema and documentation and have no file or tool access. By default a model answers in one call, with a prompt that asks for SQL, for a brief statement of ``any consequential assumption that the request and the supplied context do not settle,'' and for SQL of any other supported interpretation, without inventing alternatives. Computing is not our main question, so to test whether a model can compute an interpretation at all, we give it its best chance and let it explore. It may first run up to six rounds of five queries on the database, seeing twenty rows of each result. Exploration changes little: GPT-Sol solves 59 of 100 held-out BIRD questions without it and 61 with it, so one call is a fair default.\par}

\textbf{Statistics.} Intervals resample tasks 2,000 times, keeping pairs together; proportions use Wilson intervals; paired comparisons use exact McNemar tests; differences are mean $\pm$ standard error.

\textbf{Baselines.} We compare our method, \method{Decide-then-answer} with its option \method{Decide-then-search} (Section~\ref{sec:method}), with three research methods and two behaviors of production assistants. \method{Plain} uses the AMBROSIA authors' prompt~\cite{ambrosia2024}, \method{Three samples} runs it three times, and \method{DtP} (disambiguate-then-parse)~\cite{saparina2025disambiguate} lists the interpretations and answers each. \method{Single answer} returns one query and states its interpretation, as Snowflake's Cortex Analyst does (``We interpreted your question as\ldots''~\cite{cortexrest2025}); we use the main query of \method{Plain}. \method{Clarifying question} returns no SQL for a request it finds ambiguous and asks the user to choose, as Cortex Analyst does (``Your question is ambiguous, here are some alternatives''~\cite{cortexrest2025}) and Databricks' Genie does where its instructions say so~\cite{genie2025}. It asks wherever our first call finds a choice open, which favors it. It needs a second turn, so we test it in a separate experiment (Section~\ref{sec:methodambrosia}).

\subsection{Checks on the tests}
\label{sec:specresults}
\label{sec:resconstruction}
\label{sec:writerswap}
\begin{figure}[t]
\centering\includegraphics[width=\columnwidth]{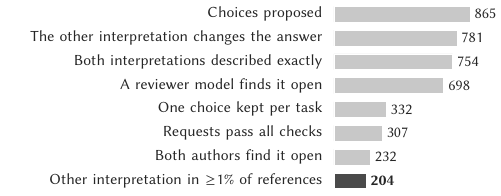}
\caption{What each construction stage keeps, from proposed choices to accepted request sets.}
\Description{Horizontal bars decreasing from proposed choices to accepted request sets.}
\label{fig:funnel}
\end{figure}

The findings rest on tests that change only the choice, so we check the tests first (Figure~\ref{fig:funnel}). Code confirms that in each of the 307 sets that pass all checks, the two queries differ in one step and the two specific requests differ only in the choice phrase. Two authors then cross-checked every set without seeing model results and kept it only if a reasonable analyst could mean either interpretation. They keep 232 of the 307 sets (76\%). Most rejected sets have an open request whose wording already favors one interpretation, or a choice that convention settles, such as ``between,'' which includes both ends in SQL and in almost every benchmark reference. We also keep only the choice types on which real references differ (Figure~\ref{fig:choices}), which leaves 204 sets on aggregation level, cutoffs, ties, counting units, missing values and one negation. Every result below uses these 204 sets. The review does not drive the results. With no human filter, models return both interpretations in only 11--48\% of the sets they can compute, so the gap holds without the review.

The generator itself is reliable. Its rules find usable queries on 152 of the 190 Spider, BIRD and LiveSQLBench databases we tried; the others lack the repeated rows the rules need. On the 12 evaluation databases it delivers 1,198 of 1,200 requested queries, all at the requested depths, and each of its rules is needed (Figure~\extref{fig:ablation}); without the look-ahead, almost no attempt reaches its target. Natural requests also lose little against exact descriptions (Figure~\extref{fig:specification}). Sonnet and GPT-Sol lose about two points, and Haiku and GPT-Luna gain. Writing from a plan keeps the three requests aligned. On 60 tasks drawn at random across choice types, it keeps the shared requirements in 52 drafts, against 32 for direct generation ($p=0.0002$), and passes all checks in 49 against 32 drafts.

\textbf{No model favors its own family.} Claude models write and review the requests and Opus judges the explanations, so a Claude model might do better on them. We repeat each of these steps with GPT-Sol, from another family, and find no such advantage. When GPT-Sol writes the natural requests of 200 random generated tasks, neither GPT-Sol nor Sonnet gains on its own text. With exploration, GPT-Sol answers 395 of the 408 specific requests that Opus wrote, as many as Sonnet (396). When GPT-Sol rewrites the open requests of 60 random sets, no model returns both interpretations significantly more or less often (all $p\ge0.18$). The Opus judge does not favor Claude models either. It credits naming the choice in 96--100\% of responses that return both interpretations, and among responses that give one interpretation it credits GPT-Luna most often (27\%).

\section{Models Make Choices Without Saying So}
\label{sec:disclosure}
A model that computes both interpretations when each is asked for usually returns only one of them for the open request, and often does not say that there was a choice. Asking for an explanation does not cause this gap. Against the same call asking only for SQL, correct answers to specific requests change by at most nine of 204 per model, and every paired interval includes zero. Nor does the explanation make models raise settled choices. On specific requests, no alternative returns the other interpretation, and at most one of 204 responses per model calls the choice open, such as Sonnet's for a request to drop districts tied with the 10th (Table~\ref{tab:scoring}). Alternatives on other choices, which the sets do not test, appear in 4--20 of 204 responses per model.

\begin{figure}[t]
\centering\includegraphics[width=\columnwidth]{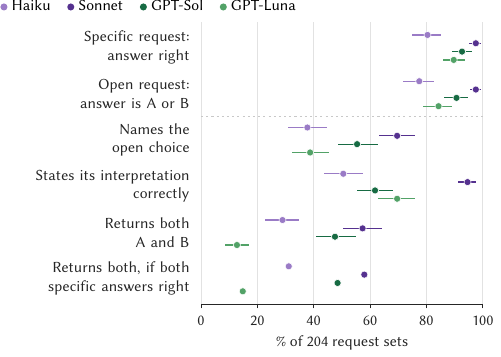}
\caption{Answering and communicating on 204 request sets, per model and measure (95\% intervals).}
\Description{Dot plot of five measures for four models with 95 percent intervals.}
\label{fig:disclosure}
\end{figure}

\textbf{Models return one interpretation where two are correct.} Returning both interpretations is much rarer than returning one correct answer (Figure~\ref{fig:disclosure}). Even among the sets in which a model answers both specific requests correctly, it returns both interpretations for the open request in only 58\% for Sonnet, 48\% for GPT-Sol, 31\% for Haiku and 15\% for GPT-Luna. Accuracy hardly separates the models, whose answers to specific requests are right in 80--98\% of sets, but reporting does. The larger model of each family returns both about two and three times as often as the smaller one, yet none closes the gap. When a model states its interpretation, the statement matches its SQL in 99--100\% of responses. The unreported interpretation matters. In 98 of the 204 sets, the two interpretations return different rows, and in the other 106 the same rows with different values, so the answer differs in every set.

\begin{figure}[t]
\centering\includegraphics[width=\columnwidth]{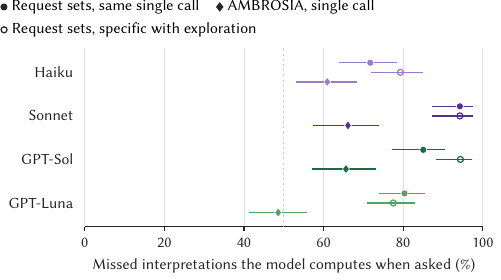}
\caption{Missed interpretations that the same model computes when asked for each of them alone, per model, on request sets with one call or after exploring the database, and on AMBROSIA (Wilson 95\% intervals).}
\Description{Dot plot with one row per model and three markers per row; most markers lie between 50 and 95 percent.}
\label{fig:decomposition}
\end{figure}

\begin{figure}[t]
\centering\includegraphics[width=\columnwidth]{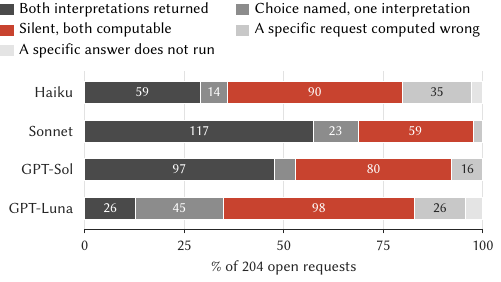}
\caption{What each model does with the 204 open requests (one call with explanation). Computable: the same call answers both specific requests correctly.}
\Description{Stacked horizontal bars, one per model, split into five outcomes; silent answers with both interpretations computable form the largest part for Haiku and GPT-Luna.}
\label{fig:outcomes}
\end{figure}

\textbf{Most misses are failures to report.} Most interpretations a model misses on the open request are ones it can compute. The same model computes the missed interpretation correctly when asked for it alone in 72--94\% of the missed sets (Figure~\ref{fig:decomposition}). Figure~\ref{fig:outcomes} sorts every open request into five outcomes, namely both interpretations returned, the choice named but one interpretation given, one interpretation given silently although the model computes both, a specific request computed wrong, and a specific answer whose SQL does not run. For Haiku and GPT-Luna the most common outcome is the silent one, in which the model can compute both interpretations but gives one without naming the choice, and for Sonnet and GPT-Sol it is the second most common. A score that only checks whether both interpretations appear among the outputs, as top-$k$ coverage~\cite{ambiqt2023} does, cannot tell these outcomes apart. Sampling does not surface the other interpretation either. On 55 randomly drawn sets, five SQL-only samples contain both interpretations in at most 14 for any model, while one explained response from Sonnet does so in 34. What is missing is a prompt to report the choice; SQL-only samples mostly repeat one interpretation.

\begin{figure}[t]
\centering\includegraphics[width=\columnwidth]{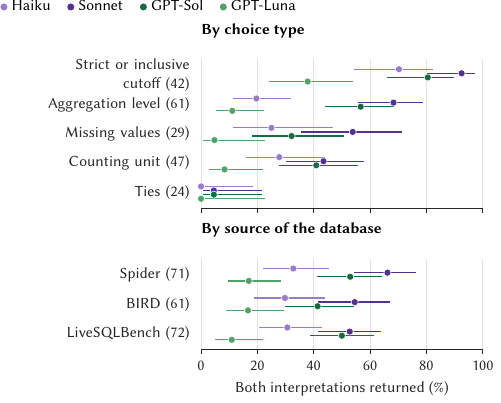}
\caption{Open requests answered with both interpretations, by choice type and by source of the database, among the sets in which the same model answers both specific requests right (accepted sets in parentheses; Wilson 95\% intervals).}
\Description{Dot plot with five choice types and four models; cutoffs are high for three models, ties and counting units are low for all.}
\label{fig:bytype}
\end{figure}

\textbf{The gap is widest for ties and counting units.} Sonnet, the best model, returns both interpretations for 68\% of aggregation-level sets but for only 43\% of counting-unit sets and one of 22 tie sets, and no model returns both for more than one tie set (Figure~\ref{fig:bytype}). These choices hide in the SQL, in a \sql{LIMIT} or a \sql{COUNT} over a join, neither of which the request mentions. Cutoffs are the exception, because most cutoff requests give a value ``as the cutoff'' with no comparison word, which invites both the strict and the inclusive interpretation. The gap holds on every source, and no model returns both interpretations for more than 67\% of the sets from any source.

\textbf{Naming a choice is not the same as giving the other interpretation.} A model that says a choice is open often stops there. Among the sets in which a model names the choice and can compute both interpretations in the same call, it still returns only one interpretation in 10--65\% of them, and GPT-Luna does so most often. Such a response warns the user but leaves them to ask for the other interpretation, a step the model could have taken itself, since it computes that interpretation when asked for it alone.

\begin{figure}[t]
\centering\includegraphics[width=\columnwidth]{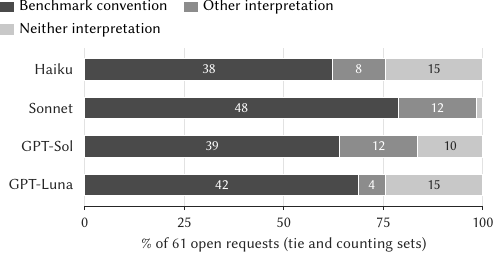}
\caption{Main answer to the open request on 61 tie and counting sets: the interpretation BIRD and Spider references take by convention, the other interpretation, or neither.}
\Description{Stacked horizontal bars per model; the conventional interpretation outnumbers the other interpretation by at least three to one for every model.}
\label{fig:conventions}
\end{figure}

\textbf{Silent choices follow conventions on our sets, but not on BIRD.}\label{sec:filterresults} On our request sets, when a model settles a choice without saying so, it usually takes the interpretation that benchmark authors take. In 61 tie and counting sets, one interpretation is the convention of BIRD and Spider references (Section~\ref{sec:defaults}), namely a plain \sql{LIMIT} rather than every row tied at the cut-off, and counting joined rows rather than distinct entities. The conventional interpretation is the intended one in 29 of these sets, so the design favors neither. Yet when a model's main answer is one of the two interpretations, it is the conventional one in 76--91\% of cases for every model (Figure~\ref{fig:conventions}). The conventional interpretation is also the simpler SQL, so these data cannot separate habit from simplicity. On BIRD's own tie and counting questions the preference reverses: more answers match the other interpretation (90) than the reference (37; Section~\ref{sec:impact}). Which interpretation a model takes silently thus varies with the request, and a user cannot tell which one an answer computes unless the model says so.

\textbf{Many misses on AMBROSIA are also failures to report.} The pattern is not an artifact of our tests. AMBROSIA pairs each ambiguous question with one unambiguous question per interpretation, and Saparina and Lapata~\cite{saparina2025disambiguate} observe that models usually give one preferred interpretation. We split each missed interpretation into one the model cannot compute and one it computes but does not report. On 120 ambiguous test questions per model, the model computes 49--66\% of the interpretations it misses when asked the unambiguous version (Figure~\ref{fig:decomposition}). Even when a model answers every unambiguous version correctly, it returns all interpretations for the ambiguous question in only 12--28\% of questions. On the 40 vagueness questions per model, which ask which column or value a phrase means, models compute 48--56\% of the interpretations they miss, so the gap also covers the referent-column choices that our request sets leave out. On both benchmarks, models leave open choices unreported, whether the ambiguity lies in wording or calculation.

\section{Benchmarks Cannot See the Gap}
\label{sec:diagnosis}
\begin{figure}[t]
\centering\includegraphics[width=\columnwidth]{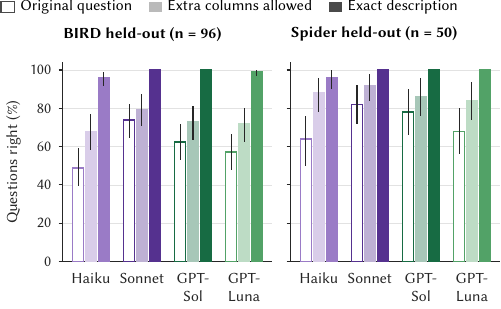}
\caption{Held-out questions: original question versus the exact description of the reference SQL (95\% intervals).}
\Description{Grouped bars for four models on BIRD and Spider held-out questions under three conditions.}
\label{fig:real}
\end{figure}

BIRD and Spider score a model against one interpretation even where a question leaves the choice open, and which interpretation they take depends on who wrote the question. On held-out questions, the four models answer 49--74\% of BIRD questions correctly as worded but 96--100\% from the exact description of the reference SQL (Figure~\ref{fig:real}). Allowing extra or reordered output columns closes only part of this gap, and on Spider the remaining gap is smaller.

\begin{figure}[t]
\centering\includegraphics[width=\columnwidth]{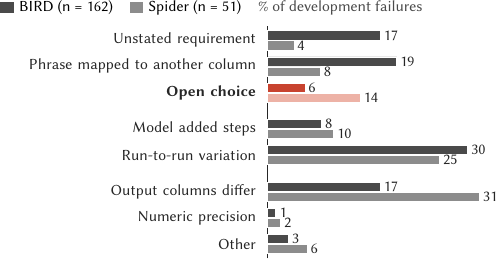}
\caption{Why the original question failed, on development questions answered right from the exact description and wrong as worded, one cause per question.}
\Description{Eight causes, one row each, with a bar for BIRD and one for Spider; open choices are shown in red.}
\label{fig:causes}
\end{figure}

The exact description gives the reference's steps, columns and choices at once, so this gap bounds what the wording costs rather than measuring ambiguity alone. To see what it consists of, Opus assigns one cause to each development question answered correctly from the description and wrongly as worded (Figure~\ref{fig:causes}). Output columns, a phrase mapped to another column or value, unstated requirements and run-to-run variation cause most failures. A second labeler from another family, GPT-Sol, given the same evidence, agrees on the cause for 53\% of the questions (Cohen's $\kappa=0.44$); it often calls Opus's run-to-run variation output shape and finds an open choice in 35 questions where Opus finds 16, so the figure, if anything, undercounts open choices. Output columns are a matter of form, as the answer holds the right values in a different shape, such as a name instead of an ID. Open choices, which our library collects, are one of these causes. Failures undercount them, because an answer following the benchmark's convention is scored right; Section~\ref{sec:impact} measures them on every question that leaves a choice open.

Other choices are about which column a word means. For ``Name schools in Riverside\ldots'', filtering the district field returns six rows, the county field sixty and the city field seven. Code cannot tell from a reference which column the question's word could also mean, so request sets leave this type to AMBROSIA's vagueness questions.

\subsection{References follow unstated conventions}
\label{sec:defaults}
\begin{figure}[t]
\centering\includegraphics[width=\columnwidth]{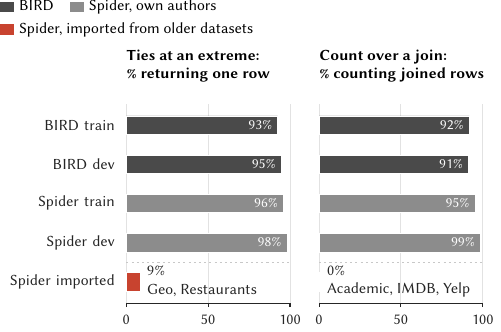}
\caption{Share of open questions whose reference takes the usual interpretation, by author group, for the two choices on which groups disagree, ties and counting.}
\Description{Two panels of horizontal bars. BIRD and Spider bars are at 91 to 99 percent; the imported Spider groups are at 9 percent (ties) and 0 percent (counting).}
\label{fig:defaults}
\end{figure}

When a question leaves a choice open, its reference still takes one interpretation, and the interpretation depends on the author group. For other common choices, such as join type, missing values and whether ``between'' includes its bounds, every author group takes the same interpretation in at least 91\% of open questions. On ties and counting, the groups disagree (Figure~\ref{fig:defaults}). Asked for an extreme value, BIRD and Spider references return one row in over 90\% of open questions, but the Geo and Restaurants questions that Spider imported from older datasets return one row in 21 of 247. Asked for a count over a join, BIRD references count joined rows in most open questions, and the imported Academic, IMDB and Yelp questions in none of 39. Each group keeps its convention in training and development data alike, so examples retrieved from a benchmark's training data can carry its authors' convention into a model's answers. For the same wording, a model that follows BIRD's convention is therefore scored right on BIRD and wrong on the imported Spider questions.

\begin{table}[t]
\caption{Six of the 17 BIRD questions on which all four models return the other interpretation, confirmed by two authors; BIRD scores each of these answers wrong.}
\label{tab:answers}
\centering\footnotesize\setlength{\tabcolsep}{3pt}\renewcommand{\arraystretch}{1.15}
\begin{tabularx}{\columnwidth}{@{}Y>{\raggedright\arraybackslash}p{1.75cm}>{\raggedright\arraybackslash}p{1.75cm}@{}}
\toprule
Question & BIRD reference & Other interpretation\\
\midrule
``Who is the tallest superhero?'' & 1 hero & 2 tied heroes\\
``\ldots\ cards in the set Coldsnap with the highest converted mana cost.'' & 1 card & 12 tied cards\\
``What are the expenses of the budget with the lowest remaining?'' & 1 expense & 2 tied expenses\\
``How many of the banned cards are white border?'' & 258 rows & 89 cards\\
``\ldots\ normal glutamic pylvic transaminase level, how many of them are male?'' & 1{,}924 lab records & 45 patients\\
``\ldots\ diagnosed with SJS, how many of them have a normal level of total protein?'' & 780 lab records & 37 patients\\
\bottomrule
\end{tabularx}
\end{table}

\subsection{The unstated interpretation changes answers}
\label{sec:impact}
\begin{figure}[t]
\centering\includegraphics[width=\columnwidth]{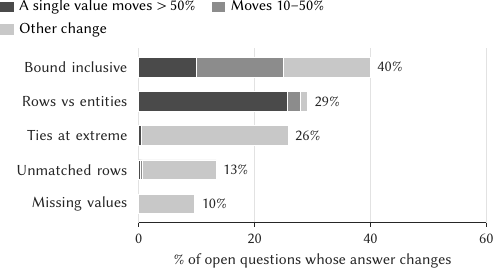}
\caption{Open BIRD development questions whose answer changes under the other interpretation, by choice type and by how far a single answer value moves.}
\Description{Stacked horizontal bars for five choice types; inclusive bounds change 40 percent of answers and counting rows versus entities 29 percent, mostly with values moving by more than half.}
\label{fig:impact}
\end{figure}

The choice a reference makes for an open question is not cosmetic. For each open BIRD development question, we apply the library's edit for the other interpretation to the reference SQL wherever it gives a valid interpretation, and run both. Of 664 such questions, 156 (23\%) change their answer (Figure~\ref{fig:impact}). These are one in ten of all BIRD development questions. Counting moves the answer most, often by more than half of its value. For ``How many accounts in North Bohemia has made a transaction with the partner's bank being AB?'', the reference counts joined rows and returns 2,260, while counting each account once returns 55. Ties change about a quarter of answers, often the list of returned items, and join type and missing-value filters change about one in ten. The data behind these choices are common. Every one of the 12 evaluation databases has at least two kinds of data on which an unstated choice changes an answer, such as joins that repeat rows, columns with missing values and maxima shared by several rows.

\textbf{BIRD scores the other interpretation as wrong.} On the 79 questions checked below, accepting both interpretations raises accuracy from 25\% to 57\%; the difference is answers that BIRD counts as wrong SQL, although the question allows them. We ask the four models 137 BIRD development questions on which a tie, counting or join edit of the reference changes the answer, and score each answer against both queries. Two authors checked that the edit gives a valid interpretation of the question, and 79 questions pass. Of the 236 answers BIRD marks wrong on them, 99 return the other interpretation. On tie and counting questions, more answers match the edited reference (90) than the reference itself (37), and on 17 questions every model returns the other interpretation (Table~\ref{tab:answers}).

\section{A Two-Phase Method for Open Choices}
\label{sec:method}
A conversation about an open choice has two moments, before the user says which interpretation they mean and after. Production assistants ask or return one interpretation before (Section~\ref{sec:intro}) and rewrite the request after (Section~\ref{sec:rewrite}). Our method has one phase for each moment, and each step follows from a finding of this paper. Models state their interpretation accurately and almost never call a settled choice open (Section~\ref{sec:disclosure}), so one call can decide whether a choice is open; they compute most interpretations they miss once each is asked for alone, so each gets its own call; and a rewrite keeps the choice but loses other requirements (Section~\ref{sec:rewrite}), so each interpretation is written as the full request and the pick is added as one clause, leaving the rest word for word.

\subsection{Phase 1: decide, then answer}
\label{sec:methoddesign}
\begin{figure}[t]
\centering
\includegraphics{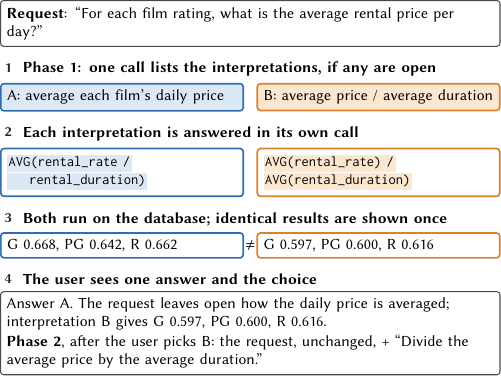}
\caption{Both phases on the film request. Blue and orange: interpretations A and B, as in Figure~\ref{fig:teaser}.}
\Description{Four steps: decide whether a choice is open and list the interpretations, answer each in its own call, run both and compare the results, show one answer with the choice and carry the user's pick as one added clause.}
\label{fig:methodsteps}
\end{figure}

Phase 1, \method{Decide-then-answer}, runs before the user replies (Figure~\ref{fig:methodsteps}). First, one call lists the interpretations of the request, each written as the full request with exactly one interpretation; if it lists only one, one more call answers it, and nothing else runs, so a clear request costs two calls and one query. Second, where a choice is open, each interpretation it lists, at most four, is answered in its own call. Third, every answer runs on the database, SQL that fails is dropped, and identical results are shown once. The user sees the first answer with the others as alternatives, and phase 2 starts when the user picks one (Section~\ref{sec:rewrite}). An option, \method{Decide-then-search}, first searches wide in the second step, with two more calls that list the interpretations and then any that are missing, using \method{DtP}'s prompts~\cite{saparina2025disambiguate}, for users who must not miss an interpretation.

\subsection{Phase 1 against products and baselines}
\label{sec:methodambrosia}
\label{sec:exchange}
We developed the method on 60 AMBROSIA questions. The final test uses 150 new test questions per model (90 ambiguous, 60 unambiguous) and 111 random request sets, with Sonnet and GPT-Sol.

\begin{figure}[t]
\centering\includegraphics[width=\columnwidth]{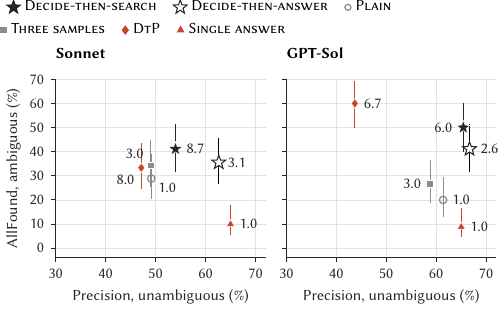}
\caption{Methods on 150 new AMBROSIA test questions per model; calls per question beside each point (95\% intervals).}
\Description{Two scatter panels, Sonnet and GPT-Sol: decide-then-search, decide-then-answer, plain, three samples, DtP and single answer.}
\label{fig:method}
\end{figure}

\begin{table}[t]
\caption{\method{Decide-then-answer}, \method{Decide-then-search} and the baselines of Section~\ref{sec:setup}, measured the same way on both datasets: coverage (all interpretations returned, of 90 ambiguous AMBROSIA questions and 111 open requests), precision on clear or specific requests (\%), and calls per request. Bold: best per model.}
\label{tab:methods}
\centering\footnotesize\setlength{\tabcolsep}{3.2pt}
\begin{tabular}{@{}llrrrrrr@{}}
\toprule
& & \multicolumn{3}{c}{AMBROSIA} & \multicolumn{3}{c}{Request sets} \\
\cmidrule(lr){3-5}\cmidrule(l){6-8}
Model & Method & Cover & Prec. & Calls & Cover & Prec. & Calls \\
\midrule
Sonnet & \method{Plain} & 26 & 49.1 & \textbf{1.0} & 64 & 94.6 & \textbf{1.0} \\
 & \method{Three samples} & 31 & 49.0 & 3.0 & 72 & 92.0 & 3.0 \\
 & \method{DtP} & 30 & 47.1 & 8.0 & 10 & 18.6 & 8.0 \\
 & \method{Single answer} & 9 & \textbf{65.0} & \textbf{1.0} & 0 & \textbf{98.2} & \textbf{1.0} \\
 & \method{Decide-then-answer} & 32 & 62.6 & 3.1 & 78 & 96.1 & 2.6 \\
 & \method{Decide-then-search} & \textbf{37} & 54.0 & 8.7 & \textbf{79} & 89.2 & 6.7 \\
\midrule
GPT-Sol & \method{Plain} & 18 & 61.4 & \textbf{1.0} & 47 & 83.3 & \textbf{1.0} \\
 & \method{Three samples} & 24 & 58.8 & 3.0 & 60 & 82.6 & 3.0 \\
 & \method{DtP} & \textbf{54} & 43.7 & 6.7 & 35 & 67.3 & 5.3 \\
 & \method{Single answer} & 8 & 65.0 & \textbf{1.0} & 0 & 82.9 & \textbf{1.0} \\
 & \method{Decide-then-answer} & 37 & \textbf{66.7} & 2.6 & 62 & \textbf{93.7} & 2.5 \\
 & \method{Decide-then-search} & 45 & 65.4 & 6.0 & \textbf{68} & 91.5 & 5.3 \\
\bottomrule
\end{tabular}
\end{table}

\textbf{More interpretations than today's answer, at no loss of precision.} Pooled over both models and paired by item, \method{Decide-then-answer} returns every interpretation for $28.9 \pm 3.5$ points more ambiguous AMBROSIA questions than \method{Single answer} and for $63.1 \pm 3.2$ points more open requests, with no loss of precision ($-0.3 \pm 3.7$ points on AMBROSIA, $+4.4 \pm 2.0$ on request sets), and its main answer is right on 6.3 points more specific requests. Precision holds because the first step finds no open choice in 86--90\% of specific requests, which then get one answer, as from \method{Single answer}. Against \method{Plain}, \method{Three samples} and \method{DtP}, it is the most precise, by 9.4 to 19.2 points on AMBROSIA and 5.9 to 52.0 points on request sets, all significant; it covers more than \method{Plain} and \method{Three samples}, and more than \method{DtP} on request sets, by 42.8 points, where \method{DtP}'s restatements lose other requirements. Only on AMBROSIA does \method{DtP} cover more, by 8.3 points, at 19.2 points less precision and nearly three times the calls, having no step for clear questions.

\textbf{The first step brings the gain.} \method{Decide-then-search} is \method{DtP} with the first step added, so their difference isolates that step: the same AllFound ($-1.1$ points) and 14.3 points more precision. The gain is not from returning fewer alternatives. With \method{DtP} cut, per question, to the number of interpretations that the first step lists, this budget finds every interpretation on $7.6 \pm 2.0$ points more ambiguous questions when spent where the first step finds a choice than at random, and \method{Decide-then-answer} still finds every interpretation on $15.0 \pm 2.9$ points more ambiguous questions at no lower precision. \method{Decide-then-search} finds every interpretation for 7.2 points more ambiguous questions but is 5.0 points less precise, at 2.3 times the calls, so it suits users who must miss no interpretation.

\textbf{Two-turn experiment: fewer user turns than asking.} \method{Clarifying question} needs a second turn, so we test it separately. A simulated user has one interpretation in mind and, when the first turn does not show it, states it in one more turn; the model then answers the specific request, the same second answer for every method. Per 100 requests, with one in five AMBROSIA questions ambiguous, \method{Decide-then-answer} needs $35.0 \pm 3.3$ fewer user turns than \method{Clarifying question} and $14.9 \pm 2.8$ fewer than \method{Single answer}, and $40.8 \pm 2.4$ and $22.1 \pm 2.0$ fewer on request sets. It also ends right on $9.1 \pm 2.2$ and $12.6 \pm 2.5$ more AMBROSIA requests and on $3.4 \pm 0.9$ and $6.1 \pm 1.1$ more request-set requests, because an answer shown in the first turn needs no second answer that can go wrong, at 1.1 to 1.5 more calls per request.

\subsection{Phase 2: keep the request, add the pick}
\label{sec:rewrite}
When the user picks an interpretation, today's systems rewrite the request. Cortex Analyst rewrites each follow-up into a self-contained question~\cite{xu2024cortex}, the Fabric data agent rephrases each question~\cite{fabric2026agent}, LangChain's ConversationalRetrievalChain~\cite{langchain2024crc} and LlamaIndex's CondenseQuestionChatEngine~\cite{llamaindex2024condense} condense each follow-up into a standalone question, and \method{DtP} writes each interpretation as a short question. Each rewrite is a new copy that can lose a requirement, so phase 2 keeps the request and adds the pick as one clause, leaving the rest unchanged.

Request sets test this, as their three requests differ only in the choice. For each interpretation of a request set, the same SQL prompt receives one of four texts: (a) the original specific request; (b) the open request rewritten by the model as one interpretation with \method{DtP}'s prompt, told the intended interpretation; (c) the open request unchanged, followed by a one-sentence description of the interpretation; (d) the open request and the interpretation condensed by the model into a standalone question with LangChain's prompt~\cite{langchain2024crc}; we report 363 interpretations drawn at random from the 204 sets, with all four models and one SQL prompt.

\begin{figure}[t]
\centering\includegraphics[width=\columnwidth]{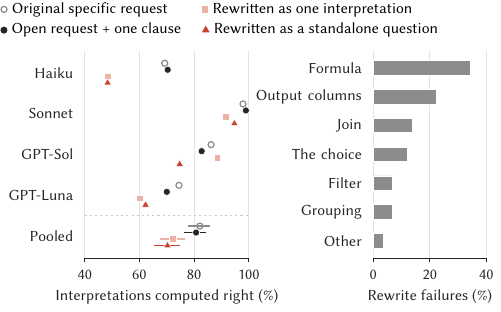}
\caption{Editing versus rewriting a request (363 interpretations). Left: interpretations computed right from four request texts. Right: what the rewrite lost, by kind.}
\Description{Left: dot plot per model of accuracy for four request texts; the two rewrites lie below the edited request. Right: horizontal bars, formula largest.}
\label{fig:rewrite}
\end{figure}

Keeping the request and adding one clause computes 80.7\% of the interpretations correctly, as many as the original specific requests ($p=0.57$, exact McNemar; Figure~\ref{fig:rewrite}). Both rewrites compute fewer, 8.5 points fewer for the interpretation rewrite ($p=0.0004$) and 10.5 for the standalone question ($p<0.0001$), and the loss is largest for Haiku and GPT-Luna. The rewrite seldom loses the choice itself. Of the interpretations answered correctly from the original request and wrongly from the rewrite, the judge finds the choice lost in only 12\%; the rest lose another requirement, most often a formula, an output column or a join (Figure~\ref{fig:rewrite}, right).

\textbf{The same holds on real questions.} We repeat the comparison on 85 BIRD development questions where one library edit of the reference SQL changes the answer. A simulated user follows each question with one sentence that settles the choice, such as ``If several rows tie at the cut-off, return all of them,'' and the edited reference is the gold answer. Over 170 question--model pairs with Sonnet and GPT-Sol, adding the clause beats the standalone question, 58.2\% against 47.6\%. As on our request sets, the rewrite keeps the clarified choice and loses something else; the gap is largest for ties.
\section{Implications}
\label{sec:lessons}
\textbf{Systems could answer open choices instead of asking.} Deciding first and answering each interpretation in its own call adds the other interpretations at no loss of precision and saves the user turns that asking costs (Section~\ref{sec:method}).

\textbf{Reporting could be a training target.} The models we test already compute both interpretations of most requests; what they mostly lack is the habit of reporting the second one. Sonnet, the strongest model, still misses it in four of ten sets where it can compute it, so training may help more than scale, and request sets could supply the signal. Each interpretation has its own executable reference, so a reward can credit an open answer only when it returns both interpretations and charge a specific answer for every alternative it adds, which rewards reporting but not guessing.

\textbf{Benchmarks could reward stating assumptions.} Current benchmarks score a single answer per question, which suits questions with one correct interpretation. For a question that leaves a choice open, a complementary practice is to let the model state the assumption it made and the alternative it set aside, and to credit both when each matches its own reference. Request sets make this possible because each interpretation has its own reference, so a benchmark can report accuracy on specific requests, reporting on open requests, and restraint on settled ones.

\section{Related Work}
\label{sec:related}
\textbf{Natural language interfaces to databases.} NaLIR~\cite{nalir2014} fixed ``easy'' ambiguities quietly and, for each ``hard'' one, explained its interpretation and offered several for the user to choose from. Its ambiguities lie mainly in mapping words to schema elements and in how phrases attach. Our results show that the alternatives NaLIR asked users to choose among are now cheap to compute; what is missing is the report. Our method applies NaLIR's interaction to calculation choices and lets execution decide which are ``hard''.

\textbf{Evaluation.} Spider tests generalization across schemas~\cite{spider2018}; BIRD adds large databases and external knowledge~\cite{bird2023}; KaggleDBQA uses realistic databases and documentation~\cite{kaggledbqa2021}; Spider 2.0 tests enterprise workflows~\cite{spider22025}. SParC and CoSQL add context and dialogue~\cite{sparc2019,cosql2019}, and BIRD-INTERACT tests clarification and recovery~\cite{birdinteract2026}. It scores whether a system completes a task after asking the user, whereas request sets score whether it reports a choice it made without asking. TrustSQL rewards a model for abstaining when it cannot answer reliably and penalizes wrong answers~\cite{lee2024trustsql}; request sets reward a different behavior, answering while reporting the open choice. Error detectors flag executable SQL that is wrong against one gold query~\cite{chen2023error}, and verifiers turn such silent failures into abstentions~\cite{alrashed2026verification}; a silent choice passes both, because its SQL is correct for one interpretation. Test suites~\cite{testsuite2020} catch wrong queries whose results match by chance; our concern is the opposite error, in which a correct query for another valid interpretation is rejected because the reference fixes one interpretation. Pourreza and Rafiei re-evaluated top models on Spider and BIRD by hand and found a perfect score out of reach ``due to the multiple interpretations that can be derived from the provided samples''~\cite{pourreza2023evaluating}, and Wretblad et al.\ found noise in BIRD's questions and references whose correction changed which method ranks first~\cite{wretblad2024noise}. We measure the first problem by construction rather than inspection and keep possible reference errors apart from open choices, because a wrong reference and a valid but different one need different fixes.

\textbf{Ambiguous requests.} The closest benchmarks are AmbiQT~\cite{ambiqt2023} and AMBROSIA~\cite{ambrosia2024} (Table~\ref{tab:comparison}). AmbiQT makes Spider questions ambiguous by changing the schema, with synonym columns or tables, split tables and precomputed aggregates, and checks whether both gold queries appear in the top five outputs. AMBROSIA has human-written questions with scope, attachment and vagueness ambiguity on generated databases. Both locate ambiguity in the schema or the wording. The choices behind the conventions of Section~\ref{sec:defaults}, such as ties, counting units, missing values and aggregation level, lie in the calculation, outside both taxonomies. Outside SQL, AmbigQA asks for every plausible answer to an ambiguous question~\cite{min2020ambigqa}, and conversational search studies when a system should ask a clarifying question~\cite{aliannejadi2019asking}.

\textbf{Generation, SQL methods and semantic layers.} DBPal~\cite{dbpal2020} and GAZP~\cite{gazp2020} generate training pairs for a target database; we also start from the database, but for evaluation. RAT-SQL, PICARD, DIN-SQL, DAIL-SQL, CHASE-SQL and Alpha-SQL~\cite{ratsql2020,picard2021,dinsql2023,dailsql2024,chasesql2025,alphasql2025} improve how the SQL is generated. LookML has developers write each measure ``once, in one place''~\cite{lookml2025}, the dbt Semantic Layer centralizes metric definitions~\cite{dbtsl2025}, and Cortex Analyst generates SQL from them~\cite{cortexdocs2025}. A layer settles the choices its authors foresaw, usually a measure's formula and counting unit; per-request choices, such as ties or the rows a ratio includes, stay open, and these are the choices our method finds and answers for the user.

\section{Conclusion}
\label{sec:discussion}
Text-to-SQL models make choices without saying so. Request sets, which change only one calculation choice, make this visible. Four models compute both interpretations of most sets when each is asked for, yet return both for the open request in only 15--58\% of those sets. Benchmarks cannot see the gap, because their references take one interpretation by conventions that differ by author group; over two in five answers BIRD marks wrong on such questions give the other interpretation. A method that decides first whether a choice is open and answers each interpretation in its own call adds what \method{Single answer} drops at no loss of precision, needs fewer user turns than asking, and keeps the request when the user picks. The models already hold the other interpretation; what is missing is a step that shows it to the user, and a benchmark that asks for it and rewards it. Until then, a plausible number from a text-to-SQL system can answer a different question than the analyst asked, and nothing in the answer says so. We release the request sets, the library of choices, every model response and the scoring code.

\textbf{Limitations and future work.} The generator covers a subset of analytical SQL, and the library comes from two benchmarks on SQLite, so other benchmarks may hold choice types we have not seen. The request sets cover every choice type on which real references differ, except quantifier, which the generator does not yet build, and referent column, which code cannot read off a reference; for the latter, AMBROSIA's vagueness questions show the same gap. Each request set tests one choice at a time, so requests that leave two choices open at once are not covered. Each open request is scored on the two interpretations that real references take; further valid ones are unscored, so the gaps are conservative (Section~\ref{sec:metrics}). The conventions are read by rules that see one SQL form each, so their counts are lower bounds. One model family wrote the request sets, but the results hold with another writer and without the human review (Section~\ref{sec:specresults}). The BIRD rescoring uses one call per question without exploration, so its shares may shift with stronger prompting. \method{Single answer} and \method{Clarifying question} are rebuilt from the products' documentation, not run on the products, and the two-turn experiment uses a simulated user who always recognizes the interpretation it means. Other SQL dialects, longer sessions and a study of how analysts act on a reported choice are our next steps.
\clearpage 
\bibliographystyle{ACM-Reference-Format}
\bibliography{references}
\clearpage
\setcounter{figure}{0}\setcounter{table}{0}\setcounter{algorithm}{0}
\renewcommand{\thefigure}{S\arabic{figure}}\renewcommand{\thetable}{S\arabic{table}}\renewcommand{\thealgorithm}{S\arabic{algorithm}}
\noindent These appendices add evidence and detail to the paper; they do not repeat it. Appendix~\ref{sec:datamap} maps the data and the experiments. Appendix~\ref{sec:appendixlibrary} shows that the choices are common: how often benchmark references meet each type, and on how many databases it occurs. Appendix~\ref{sec:cases} lists documented cases outside SQL in which one such choice changed a consequential number. Appendix~\ref{sec:filtersets} reports the request sets whose choice lies in a filter condition. Appendices~\ref{sec:querydetails} and~\ref{sec:furtherdetails} give the generator's details, the seeds and the versions. Appendix~\ref{sec:typeexamples} shows examples of every type, selected by a fixed rule. The data and code behind every number are in the reproducibility package of the repository.
\appendix
\section{Data and Experiments}
\label{sec:datamap}
\begin{figure*}[t]
\centering\includegraphics{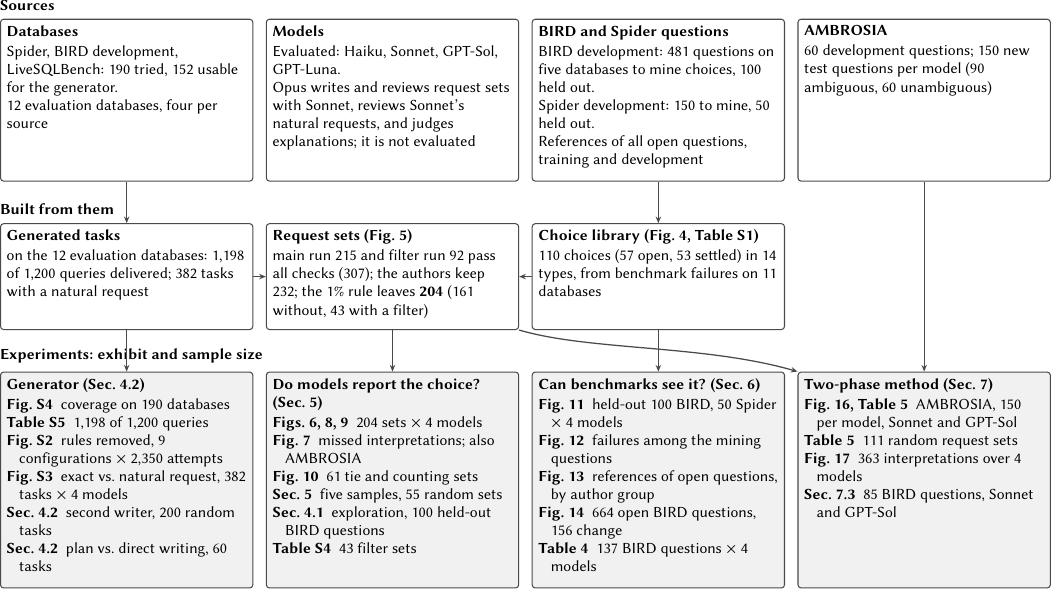}
\caption{Data and experiments: sources (top), data built from them (middle), and the experiments with their sample sizes (bottom). Arrows show what is built from what and what each experiment uses.}
\Description{Three rows of boxes. Sources: databases, models, BIRD and Spider questions, AMBROSIA. Built data: generated tasks, request sets, choice library. Experiments: generator, reporting, benchmarks, two-phase method, each with sample sizes.}
\label{fig:datamap}
\end{figure*}

Figure~\ref{fig:datamap} shows the data of the paper at three levels: the sources, the data built from them, and each experiment with its sample size. Request sets are built on the evaluation databases from generated tasks and from the choice library, which is mined from BIRD and Spider development questions; the held-out questions are not used for mining. Three subsets come from random draws made before the final selection of request sets, and the figure gives each draw with the part of it that lies in the final sets. Each experiment lists the exhibits it supports, so every figure and table of the paper can be traced to its data. The counts are generated from the same data files as the paper's numbers.

\section{The Choices Are Common}
\label{sec:appendixlibrary}
\label{sec:catalogue}
The choices in the library are not quirks of one benchmark or one database. Table~\ref{tab:librarydetail} lists every type, how often BIRD and Spider references meet it, and how many request sets test it. Table~\ref{tab:catbreadth} shows on how many databases, and in which sources, each type occurs. Appendix~\ref{sec:typeexamples} gives examples of every type that a reader can check by hand.

\paragraph{The library.} Table~\ref{tab:librarydetail} gives the labels and counts in the fixed example library for every type. ``Open'' includes both unresolved and contested records. ``Settled'' denotes failure records the reviewers label as settled by the wording; this label is not the word-list check used for the open-question counts of Section~\ref{sec:defaults}. Failures that differ only in output format (51 with two accepted interpretations, 10 with one) are not choices of meaning and are not in the table. Possible reference errors (27) and unplaced cases (four) are also kept apart. For Table~\ref{tab:comparison}, AmbiQT's column and table ambiguity and AMBROSIA's vagueness map to \emph{referent column}, AMBROSIA's scope to \emph{quantifier}, and attachment to \emph{condition linkage}.

\begin{table}[t]
\caption{All choice types (Figure~\ref{fig:choices} shows the major ones). Library: choices of meaning in benchmark failures, open (the wording fits both interpretations) or settled (it selects one). References: open BIRD and Spider questions, by the interpretation their reference SQL takes: the usual one, or the other with its share (\%), after the less common interpretations were reviewed; --: not readable from the SQL. Sets: the 204 accepted request sets, of the types in which at least 1\% of open questions have a reference that takes the other interpretation. The library's value-representation type is split by its SQL edit: comparison strictness is a strict cutoff, any other edit value matching.}
\label{tab:librarydetail}
\centering\footnotesize\setlength{\tabcolsep}{2pt}
\begin{tabular*}{\columnwidth}{@{}l@{\hspace{3pt}}l@{\extracolsep{\fill}}rrrrr@{}}
\toprule
& & \multicolumn{2}{c}{Library} & \multicolumn{2}{c}{References} & \\
\cmidrule{3-4}\cmidrule{5-6}
Choice type & The choice & Open & Settled & Usual & Other (\%) & Sets\\
\midrule
Aggregation level & Rows or group averages? & 6 & 2 & 244 & 15 (5.8) & 61\\
Calculation & Which formula is meant? & 1 & 1 & 613 & 1 (0.2) & 0\\
Quantifier & ``Than any'': one or all? & 3 & 0 & 8 & 7 (46.7) & 0\\
Condition linkage & Conditions on one row? & 1 & 0 & 260 & 0 (0.0) & 0\\
Counting unit & Count rows or entities? & 7 & 27 & 2{,}521 & 237 (8.6) & 47\\
Missing values & Keep or skip missing? & 8 & 3 & 1{,}630 & 26 (1.6) & 29\\
Negation & Some non-match, or none? & 0 & 1 & 272 & 3 (1.1) & 1\\
Population scope & Keep unmatched rows? & 0 & 1 & 12{,}462 & 5 (0.0) & 0\\
Referent column & Which column is meant? & 19 & 6 & -- & -- & 0\\
Relationship role & Which link to follow? & 2 & 4 & 15 & 0 (0.0) & 0\\
Strict cutoff & Strict or inclusive? & 1 & 0 & 309 & 5 (1.6) & 42\\
Ties and selection & One tied row, or all? & 7 & 3 & 2{,}826 & 397 (12.3) & 24\\
Time scope & Period? Ends included? & 1 & 2 & 254 & 1 (0.4) & 0\\
Value matching & Exact or loose match? & 1 & 3 & 9{,}330 & 22 (0.2) & 0\\
\midrule
\multicolumn{2}{@{}l}{Total} & 57 & 53 & & & 204\\
\bottomrule
\end{tabular*}
\end{table}

\paragraph{How references are counted.} The References columns of Table~\ref{tab:librarydetail} read each reference query with an SQL parser and the schema's keys and column types, without a model, and count a question only when its wording has no cue for the choice. Ties, counting unit, missing values, population scope, ``between'' and value matching follow Section~\ref{sec:defaults}; calculation is a question that says percent, answered times 100 or as a fraction. \emph{Quantifier}: a question that says ``than any'', answered by comparing with every value (\sql{> MAX}, \sql{< MIN}, \sql{ALL}) or with at least one (\sql{> MIN}, \sql{< MAX}, \sql{ANY}). \emph{Negation}: a negated question answered with no matching record (\sql{NOT IN}, \sql{NOT EXISTS} or \sql{EXCEPT} over another table) or with some non-matching record (\sql{<>} on the many side of a join). \emph{Aggregation level}: an entity ranked by a numeric column on the many side of a join, by one row's value or by a sum or average per group. \emph{Condition linkage}: two or more conditions on different columns of the many side of a join, on one row or on separate rows (the table joined twice by the same key, or an \sql{INTERSECT}). \emph{Relationship role}: two tables that share a parent and are also connected by a link table, joined through the parent or through the link table. Which column a phrase means depends on the words, not the SQL, so the referent column has no count.

\paragraph{A lower bound.} Each rule reads one SQL form of its choice, so it detects only part of the choices mined from failures: applied to the 85 library records of the twelve types, the rules find the choice in 50, including all ties and quantifier records, 27 of 34 counting records and 4 of 6 relationship records, but none of the aggregation, time scope or value matching records, whose forms (grouping by a non-key attribute, an average used as a threshold) the rules do not read. The counts are therefore a lower bound on how often the choices arise.

\paragraph{Review of the less common interpretations.} Every reference that takes the less common interpretation was then read against fixed criteria (the SQL takes that interpretation, the wording leaves the choice open, and the other interpretation could change the answer), and every verdict is listed in the package for the authors' review. Of 1,259 such references, 540 are dropped, most often because the wording settles the choice (``list the students with the highest'', ``how many different''), the join cannot repeat the counted entity, or a \sql{LIKE} only escapes an apostrophe. Only this side was reviewed, so the shares in the table are conservative. The same drops apply to Section~\ref{sec:defaults}.

\paragraph{Breadth.} Table~\ref{tab:catbreadth} counts every library choice by the benchmark whose failure revealed it, and every accepted request set by the source of its database. The library reflects where failures were diagnosed: its choices come from eleven databases, and several types were found in one benchmark only. The request sets show that the major types are not tied to those databases. Each of the five types with request sets beyond a single one occurs on at least eight of the twelve databases, and in databases from Spider, BIRD and LiveSQLBench.

\begin{table}[t]
\caption{Top: every library choice (Table~\ref{tab:librarydetail}), by the benchmark whose failures revealed it. Bottom: every accepted request set, by the source of its database. DBs: distinct databases.}
\label{tab:catbreadth}
\centering\footnotesize
\setlength{\tabcolsep}{2pt}
\begin{tabular*}{\columnwidth}{@{}l@{\extracolsep{\fill}}rrl@{}}
\toprule
Library choice type & Choices & DBs & Found in\\
\midrule
Aggregation level & 8 & 4 & BIRD\\
Calculation & 2 & 2 & BIRD\\
Condition linkage & 1 & 1 & BIRD\\
Counting unit & 34 & 4 & BIRD, Spider\\
Missing values & 11 & 5 & BIRD, Spider\\
Negation & 1 & 1 & BIRD\\
Population scope & 1 & 1 & BIRD\\
Quantifier & 3 & 1 & Spider\\
Referent column & 25 & 4 & BIRD, Spider\\
Relationship role & 6 & 2 & BIRD, Spider\\
Strict cutoff & 1 & 1 & BIRD\\
Ties and selection & 10 & 6 & BIRD, Spider\\
Time scope & 3 & 2 & BIRD\\
Value matching & 4 & 3 & BIRD, Spider\\
\midrule
Total & 110 & 11 & BIRD, Spider\\
\bottomrule
\end{tabular*}

\vspace{6pt}

\begin{tabular*}{\columnwidth}{@{}l@{\extracolsep{\fill}}rrrrr@{}}
\toprule
Request set type & Sets & DBs & Spider & BIRD & LiveSQL\\
\midrule
Aggregation level & 61 & 11 & 20 & 16 & 25\\
Counting unit & 47 & 10 & 16 & 5 & 26\\
Missing values & 29 & 9 & 12 & 16 & 1\\
Negation & 1 & 1 & 1 & 0 & 0\\
Strict cutoff & 42 & 10 & 17 & 16 & 9\\
Ties and selection & 24 & 10 & 5 & 8 & 11\\
\midrule
Total & 204 & 12 & 71 & 61 & 72\\
\bottomrule
\end{tabular*}

\end{table}

\paragraph{Request sets.} Table~\ref{tab:librarydetail} counts the 204 accepted request sets by type, 43 of them with a choice in a filter (Appendix~\ref{sec:filtersets}). Over the 215 sets whose choice a model proposes, aggregation level contributes 63, ties 44, missing values 40, counting unit 35, calculation 15, relationship role nine, and time scope two. The remaining seven belong to the six additional labels: running-total interpretation (two), active-status filtering, comparison baseline, baseline population, baseline definition, and row-count versus quantity measurement (one each). These counts reflect the generation and selection procedure.

\section{Calculation Choices with Real Consequences}
\label{sec:cases}
Table~\ref{tab:cases} surveys documented cases, across six domains, in which one open or mistaken choice of the kind our library records changed a consequential number or decision. None is a text-to-SQL failure. Each entry was checked against the original paper or official report; we left out cases whose primary source we could not reach, such as a bank's value-at-risk spreadsheet and a classic study of graduate admissions.

\begin{table*}[t]
\caption{Documented cases where one calculation choice changed a consequential number or decision, by domain, with the closest choice type in our library.}
\label{tab:cases}
\centering\footnotesize\setlength{\tabcolsep}{4pt}\renewcommand{\arraystretch}{1.1}
\begin{tabularx}{\textwidth}{@{}>{\raggedright\arraybackslash}p{2.2cm}>{\raggedright\arraybackslash}p{3.3cm}>{\raggedright\arraybackslash}X>{\raggedright\arraybackslash}p{2.6cm}@{}}
\toprule
Domain & Case & Choice and consequence & Library type\\
\midrule
Economics and policy & Public debt and growth~\cite{herndon2014debt} & Growth was averaged within each country before averaging across countries, so one year of New Zealand data weighed as much as decades of another country. With corrected data and weighting, high-debt countries grew 2.2\% a year, not $-0.1\%$. & aggregation level\\
\specialrule{0.2pt}{1.5pt}{1.5pt}
Health care & Risk algorithm for extra care~\cite{obermeyer2019bias} & ``Need'' was computed as health-care cost. The algorithm looked accurate by standard measures, yet fixing the proxy would raise the share of Black patients receiving extra help from 17.7\% to 46.5\%. & referent column\\
\specialrule{0.2pt}{1.5pt}{1.5pt}
Health care & Hospital mortality rates~\cite{shahian2010mortality} & Four methods, given the same 2.5 million discharges, included 28--95\% of them. 12 of 28 hospitals with higher-than-expected mortality by one method had lower-than-expected mortality by another. & population scope\\
\specialrule{0.2pt}{1.5pt}{1.5pt}
Public health & 1995 UK contraceptive pill warning~\cite{furedi1999pill} & A warning of higher thrombosis risk for two pill types was followed by pill use among under-16s falling from 40\% to 27\%, and by estimated NHS costs of \pounds 21 million for maternity care and \pounds 46 million for abortions. The author concludes that risk should be more carefully presented. & how a quantity is stated (not in our library)\\
\specialrule{0.2pt}{1.5pt}{1.5pt}
Social science & Referees and skin tone~\cite{silberzahn2018many} & 29 teams analysed the same data. Their odds ratios ranged from 0.89 to 2.93; 20 teams found a significant effect and 9 did not, using 21 combinations of covariates. & several analysis choices\\
\specialrule{0.2pt}{1.5pt}{1.5pt}
Science & Brain imaging~\cite{botviniknezer2020variability} & 70 teams analysed one dataset to test the same 9 hypotheses. No two chose the same workflow, and the results of the hypothesis tests varied substantially. & several analysis choices\\
\specialrule{0.2pt}{1.5pt}{1.5pt}
Science & Gene names in spreadsheets~\cite{ziemann2016gene} & Spreadsheet software converts gene names to dates and numbers by default; about one fifth of papers with supplementary gene lists in spreadsheets contain such conversions. & value matching\\
\specialrule{0.2pt}{1.5pt}{1.5pt}
Engineering & Mars Climate Orbiter~\cite{nasa1999mco} & Thruster data in English units were used where the trajectory software expected metric units; the spacecraft was lost. & value matching\\
\specialrule{0.2pt}{1.5pt}{1.5pt}
Engineering & Ariane 5 flight 501~\cite{lions1996ariane} & A 64-bit floating-point value was converted to a 16-bit signed integer that could not hold it; the launcher broke up about 40 seconds after lift-off. & value matching\\
\bottomrule
\end{tabularx}
\end{table*}

\section{Request Sets with Filter Conditions}
\label{sec:filtersets}
The filter sets of Section~\ref{sec:filters} come from one optional rule of the generator. It adds a single condition to the first stage of a query, on the rows of that stage's table, and builds the second interpretation at the same time. The five forms are: a date range, inclusive (\sql{>=} and \sql{<=}) or exclusive (\sql{>} and \sql{<}), with ends at the 30th and 70th percentile of a date column; a cutoff at a frequent value of a measure, \sql{>= v} or \sql{> v}; ``without'' a category, no record of the entity has it (\sql{NOT IN}) or some record lacks it (\sql{IN} with \sql{<>}); two categories, on one record or on any records of the entity; and a per-entity count, over entities with matching rows (inner join) or over all entities (outer join, zero counts kept). The generator prefers forms it has used less often. The seed is 20261016; 161 of 180 requested queries are delivered, 121 of them have interpretations with different answers, and 92 request sets pass all checks.

\paragraph{Results.} Table~\ref{tab:filtersets} gives the results per model on the 43 accepted filter sets. What models raise follows the wording of the open request. Forty of the 42 cutoff requests give a value ``as the cutoff'' or ``threshold,'' with no comparison word, and the other two say ``upwards of''; models return both the strict and the inclusive interpretation in 115 of the 159 cases where they compute both. The settled request is never called open. GPT-Luna returns both interpretations least often, as on the other sets.

\begin{table}[t]
\caption{Filter request sets (43 accepted), per model: both specific requests right (with exploration); among those, open requests answered with both interpretations; open requests that name the choice (Opus judge); specific requests right (single call); missed interpretations the model computes when asked, of all missed.}
\label{tab:filtersets}
\centering\footnotesize\setlength{\tabcolsep}{4pt}
\begin{tabular}{@{}lrrrrr@{}}
\toprule
Model & Both right & Both returned & Named & Right & Missed, computable\\
\midrule
Haiku & 42 & 33 & 40 & 40 & 9 / 10\\
Sonnet & 41 & 36 & 38 & 42 & 5 / 6\\
GPT-Sol & 42 & 32 & 35 & 40 & 10 / 10\\
GPT-Luna & 38 & 14 & 25 & 41 & 24 / 29\\
\bottomrule
\end{tabular}
\end{table}

\paragraph{Rejected sets.} The authors reject all 17 date-range sets. Every date-range request says ``between,'' which SQL's \sql{BETWEEN} includes, and 254 of the 255 BIRD and Spider references whose question says ``between'' with no other cue include both ends. Among the BIRD and Spider references whose question says ``between'' with no other cue, the one that excludes the ends is a Spider reference that applies strict bounds to the start and end columns of a staff role. A BIRD reference whose evidence says \sql{BETWEEN 121 AND 130} but whose query excludes both ends is not counted, since its evidence settles the ends. Four further references use strict bounds moved one step out, or a negated range, and select the inclusive range. Of the 49 rejected sets, 17 are the date ranges, 15 are cutoffs whose wording (``passes'', ``clears'', ``on the high side of'') already makes the comparison strict, five are ``without'' requests that already mean no such record, and one is a per-entity average whose request restricts the population. The other 11 test per-entity counts with or without unmatched entities (eight) and conditions on one or any records (three), choices whose other interpretation real references almost never take (Figure~\ref{fig:choices}).

\section{Query Generation Details}
\label{sec:querydetails}
\begin{figure}[t]
\centering\includegraphics[width=\columnwidth]{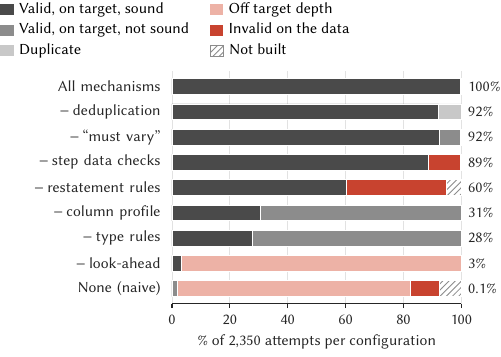}
\caption{Generation outcomes with each construction rule removed (2,350 attempts). The labels give the share valid, on target and sound.}
\Description{Stacked bars of generation outcomes for nine configurations.}
\label{fig:ablation}
\end{figure}

\begin{figure}[t]
\centering\includegraphics[width=\columnwidth]{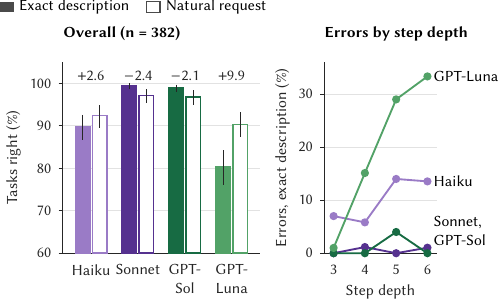}
\caption{Exact description versus natural request on 382 generated tasks. Left: accuracy; numbers give the difference in points. Right: errors from the exact description by step depth.}
\Description{Bars for four models under two request forms, and lines of error rate by step depth.}
\label{fig:specification}
\end{figure}

\begin{figure}[t]
\centering\includegraphics[width=\columnwidth]{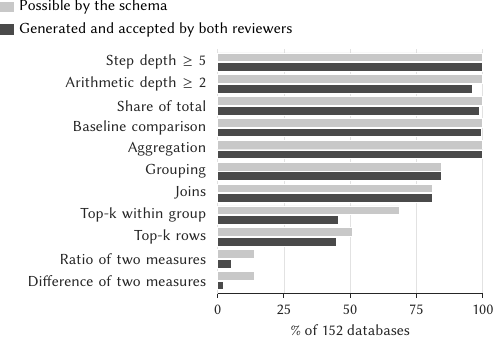}
\caption{Features the generator can build, as a share of 152 databases: possible by the schema, and generated and accepted.}
\Description{Paired horizontal bars for eleven features: possible by the schema versus generated and accepted.}
\label{fig:coverage}
\end{figure}

Algorithm~\ref{alg:families} lists the construction of a request set, Figure~\ref{fig:ablation} shows what each construction rule contributes, and Figure~\ref{fig:specification} compares the exact description with the natural request. Figure~\ref{fig:coverage} shows, for each feature, the share of 152 databases whose schema allows it and the share where a generated query with it was accepted by both model reviewers; and Table~\ref{tab:construction} the queries delivered on the 12 evaluation databases.

\textbf{Formal statement.} Let $\mathcal L(K_D)$ be the queries that the catalogue $K_D$ of database $D$ and the type rules allow, $\delta(p)$ a query's arithmetic and step depth, and $O(p)$ its operations. For target depths $d$ and required operations $R$ the generator samples from
\begin{equation}
\mathcal P(D,d,R)=\{p\in\mathcal L(K_D):\delta(p)=d,\ R\subseteq O(p),\ I_D(p)=1\},
\end{equation}
where $I_D(p)=1$ means that all data checks pass. It keeps a partial query $z$ only if some completion can still reach the target,
\begin{equation}
\operatorname{can}(z,d,R)=\mathbf 1\{\exists(d,U)\in\operatorname{Reach}(z):R\subseteq U\},
\label{eq:reach}
\end{equation}
where $\operatorname{Reach}(z)$ lists the depths and operation sets still possible. The look-ahead is exact for the rules, and no query is drawn twice. For a request set $F$ with specific requests $q_A,q_B$ and open request $q_\circ$, let $H^{(r)}_{ij}=1$ when reviewer $r$ accepts request $i$ for interpretation $j$, $f_r=1$ when $r$ finds no changed or missing requirement, and $c(F)=1$ when all code checks pass. The set is kept when
\begin{equation}
\begin{split}
\operatorname{keep}(F)=c(F)&\land\bigwedge_r\left[f_r\land H^{(r)}_{\{A,B\},\{A,B\}}=I_2\right]\\
&\land\bigvee_r\left[H^{(r)}_{\circ A}\land H^{(r)}_{\circ B}\right].
\end{split}
\end{equation}

\begin{algorithm}[t]
\caption{Build a request set around one calculation choice}
\label{alg:families}
\begin{algorithmic}[1]
\Require Databases $D$, target depths $d$, required operations $R$
\Ensure Accepted request sets and reasons for rejected sets
\For{each requested task}
\State Sample $p_A\in\mathcal P(D,d,R)$ with the look-ahead of Eq.~\eqref{eq:reach}
\State If no query is delivered, record the shortfall and continue
\State Describe $p_A$; write and check its request
\State Propose edits $p_A\rightarrow p_B$ that express a different choice
\State Execute and describe both queries; require different answers
\State Repair failed proposals once; review both interpretations
\EndFor
\State Select one choice per task at most, balancing types and databases
\For{each selected choice}
\State List what stays fixed and what changes
\State Write requests specifying A, specifying B, and leaving it open
\State Check requirements and all six request--interpretation pairs
\State Revise a failing set once and recheck
\State Keep $F$ if $\operatorname{keep}(F)$; otherwise record the reason
\EndFor
\end{algorithmic}
\end{algorithm}

\begin{table}[t]
\caption{Program construction on 12 databases, with no model calls after catalogue preparation. Every delivered SQL is distinct, has the requested depth, and passes both integrity checks.}
\label{tab:construction}
\centering\small
\renewcommand{\arraystretch}{1.12}
\begin{tabular}{@{}lrrrr@{}}
\toprule
Source & DBs & Delivered & Constructions & Seconds\\
\midrule
Spider & 4 & 398/400 & 289 & 127\\
BIRD & 4 & 400/400 & 270 & 274\\
LiveSQLBench & 4 & 400/400 & 309 & 37\\
\midrule
Total & 12 & 1198/1200 & 868 & 437\\
\bottomrule
\end{tabular}
\par\smallskip\footnotesize Source-level times are rounded individually; the total is rounded from the unrounded sum. Separate task review delivers 384/400 tasks, and language generation supplies 382/384 requests.
\end{table}
\section{Seeds, Versions and Selection Rules}
\label{sec:furtherdetails}
\paragraph{Seeds, versions and reruns.}
The grammar is \texttt{compose-v3.4}. Task requests, Spider splitting, request set selection, explanation allocation, and bootstrap resampling use seed 20261003; BIRD holdout selection uses 20261001; the paired wording comparison uses 20261004. Coverage uses seed 1 and depth-control runs seeds 1--3. Python was 3.11.5 and SQLite 3.41.2. During the experiments, the Claude command-line client changed from 2.1.287 to 2.1.289 and the Codex client from 0.147 to 0.160. Temporary usage-limit outputs were removed and retried rather than treated as SQL answers. A final query that failed under concurrent execution was rerun alone: ten such answers then matched their reference (nine GPT-Luna, one GPT-Sol). Token counts are unavailable for the explanation runs, so cost is reported as calls per request.

\paragraph{Rewrite comparison.}
The comparison of the original specific request (a) with the other texts was fixed before the run. The comparisons with the edited request (c) were added after an interim look at half of the interpretations.

\paragraph{Response selection.}
The explanation examples are the first saved responses displaying each tested pattern in file order. The worked aggregation example uses the smallest case-identifier hash among eligible request sets. Library examples use the first eligible record per type at high confidence, or medium confidence if none is high, with fixed benchmark and task ordering. Generated examples outside library labels enumerate retained cases before selecting the four shown in the main text. These selection rules support traceability; illustrative examples are not frequency estimates.

\section{Examples of Each Choice Type}
\label{sec:typeexamples}
Tables~\ref{tab:catlibrary} and~\ref{tab:catsets} show examples of each type, selected by a fixed rule rather than by hand. Library choices are taken from different databases, open choices before choices the wording settles, one from each benchmark when the type has both, shortest question first. Request sets are taken from different databases, one per source in turn, shortest open request first. Each example's question or request is given as the model received it. Answers says how the two reference queries' results differ on the database. Models gives each model's answer to the open request in one call with explanation, in the order Haiku, Sonnet, GPT-Sol, GPT-Luna: B, both interpretations returned; N, one interpretation, with the choice named; S, one interpretation given silently, although the model computes each interpretation when asked; a dash, one interpretation, and the model does not compute both. The package holds every choice and every set, with its SQL.

\onecolumn
{\scriptsize\setlength{\tabcolsep}{3pt}\renewcommand{\arraystretch}{1.05}
\begin{xltabular}{\textwidth}{@{}>{\raggedright\arraybackslash}p{2.3cm}>{\hsize=1.15\hsize\raggedright\arraybackslash}X>{\hsize=0.85\hsize\raggedright\arraybackslash}X>{\raggedright\arraybackslash}p{1.2cm}@{}}
\caption{Selected library choices, up to 3 per type.}\label{tab:catlibrary}\\
\toprule
Source & Question & The choice & Wording\\
\midrule
\endfirsthead
\toprule
Source & Question & The choice & Wording\\
\midrule
\endhead
\bottomrule
\endfoot
\multicolumn{4}{@{}l}{\textbf{Aggregation level} (3 of 8 shown)}\\[1pt]
BIRD, \texttt{california\_\allowbreak{}schools} & Which active district has the highest average score in Reading? & Whether to rank districts by one school's AvgScrRead value or by the average of AvgScrRead across each district's schools & open\\[2pt]
BIRD, \texttt{financial} & Who are the account holder identification numbers whose who have transactions on the credit card with the amount is less than the average, in 1998? Evidence: Operation = 'VYBER KARTOU' refers to credit card withdrawal & Which transactions the 'average' amount threshold is computed over & open\\[2pt]
BIRD, \texttt{thrombosis\_\allowbreak{}prediction} & For all patient born in 1982, state if their albumin is within normal range. Evidence: Year(Birthday) = '1982'; albumin is within normal range refers to ALB between 3.5 and 5.5 & Whether the normal/abnormal status is given once per laboratory record or once per patient born in 1982, since each patient can have many Laboratory rows & open\\[2pt]
\midrule
\multicolumn{4}{@{}l}{\textbf{Calculation} (2 of 2 shown)}\\[1pt]
BIRD, \texttt{financial} & Provide the IDs and age of the client with high level credit card, which is eligible for loans. Evidence: the credit card is high-level refers to card.type = 'gold'; eligible for loans refers to disp.type = 'OWNER' & How to compute the client's age from birth\_date: subtract the birth year from the current year, or count completed years of life. & open\\[2pt]
BIRD, \texttt{thrombosis\_\allowbreak{}prediction} & For all patients who are older than 55 years old, what is the percentage of female who has abnormal prothrombin time (PT)? Evidence: older than 55 years old = SUBTRACT(year(current\_timestamp), year(Birthday)) \textgreater{} 55; abnormal prothrombin time (PT) refers to PT \textgreater{} = 14; percentage = DIVIDE(SUM(PT \textgreater{} = 14 AND SEX = 'F'), SUM(PT \textgreater{} = 14)) * 100; female refers to sex = 'F'; & Which population the percentage is taken over (the denominator) & settles it\\[2pt]
\midrule
\multicolumn{4}{@{}l}{\textbf{Condition linkage} (1 of 1 shown)}\\[1pt]
BIRD, \texttt{toxicology} & Calculate the percentage of carcinogenic molecules with triple bonded Hidrogen atoms. Evidence: hydrogen refers to element = 'h'; label = '+' mean molecules are carcinogenic; triple bond refers to bond\_type = '\#'; percentage = DIVIDE(SUM(label = '+'), COUNT(molecule\_id)) * 100.0 where element = 'h' AND bond\_type = '\#'; & How a hydrogen atom is tied to a triple bond: either the hydrogen is one of the two atoms in the '\#' bond (linked through \texttt{connected}), or the hydrogen and some '\#' bond just appear in the same molecule (linked only through molecule\_id). & open\\[2pt]
\midrule
\multicolumn{4}{@{}l}{\textbf{Counting unit} (3 of 34 shown)}\\[1pt]
BIRD, \texttt{thrombosis\_\allowbreak{}prediction} & What is the average age of the male patient with high cholesterol? Evidence: average age = DIVIDE(SUM(SUBTRACT(YEAR(NOW()), YEAR(birthday))), COUNT(ID)); male patient refers to sex = 'M'; high cholesterol refers to \texttt{T-CHO} \textgreater{} = 250; & Whether the average is taken once per patient or once per matching lab record, since a patient can have several Laboratory rows with T-CHO \textgreater{}= 250 & open\\[2pt]
BIRD, \texttt{toxicology} & What is the least common element of all carcinogenic molecules? Evidence: label = '+' mean molecules are carcinogenic & What unit 'least common' is counted in: the number of carcinogenic molecules that contain the element, or the number of atoms of that element across all carcinogenic molecules & open\\[2pt]
Spider, \texttt{student\_\allowbreak{}transcripts\_\allowbreak{}tracking} & How many different degrees are offered? & What counts as one degree: a distinct degree name (degree\_summary\_name) or each degree program row in Degree\_Programs & open\\[2pt]
\midrule
\multicolumn{4}{@{}l}{\textbf{Missing values} (3 of 11 shown)}\\[1pt]
BIRD, \texttt{california\_\allowbreak{}schools} & What is the phone number of the school that has the highest average score in Math? & How to handle the top AvgScrMath row in satscores when it has no matching record in schools: drop it and take the best-scoring school that does have a record, or keep it as the answer even though its phone is missing. & open\\[2pt]
Spider, \texttt{cre\_\allowbreak{}Doc\_\allowbreak{}Template\_\allowbreak{}Mgt} & Show all template ids and number of documents using each template. & Whether templates that no document uses should appear in the output with a count of 0, or only templates that at least one document uses should be listed & open\\[2pt]
Spider, \texttt{network\_\allowbreak{}1} & Show the student IDs and numbers of friends corresponding to each. & Whether students who have no rows in Friend should be listed with a count of 0, or left out. & open\\[2pt]
\midrule
\multicolumn{4}{@{}l}{\textbf{Negation} (1 of 1 shown)}\\[1pt]
BIRD, \texttt{toxicology} & What percentage of carcinogenic-type molecules does not contain fluorine? Evidence: label = '+' mean molecules are carcinogenic; contain fluorine refers to element = 'f'; percentage = DIVIDE(SUM(element = 'f') * 100, COUNT(molecule\_id)) where label = '+'; Should consider the distinct atoms when counting; & What it means for a carcinogenic molecule to 'not contain fluorine': the molecule has at least one atom that is not fluorine, or the molecule has no fluorine atom at all. & settles it\\[2pt]
\midrule
\multicolumn{4}{@{}l}{\textbf{Population scope} (1 of 1 shown)}\\[1pt]
BIRD, \texttt{thrombosis\_\allowbreak{}prediction} & How many patients with an Ig G higher than normal? Evidence: Ig G higher than normal refers to IGG \textgreater{}= 2000; Should consider DISTINCT in the final result; & Whether the patient count only includes patients who also have records in the Examination table (because of the inner join to Examination), or every patient with any lab IgG result of at least 2000 & settles it\\[2pt]
\midrule
\multicolumn{4}{@{}l}{\textbf{Quantifier} (1 of 3 shown)}\\[1pt]
Spider, \texttt{world\_\allowbreak{}1} & What are the countries that have greater surface area than any country in Europe? & Whether 'greater than any country in Europe' means larger than at least one European country (compare to the smallest European area) or larger than every European country (compare to the largest European area) & open\\[2pt]
\midrule
\multicolumn{4}{@{}l}{\textbf{Referent column} (3 of 25 shown)}\\[1pt]
BIRD, \texttt{california\_\allowbreak{}schools} & Please specify all of the schools and their related mailing zip codes that are under Avetik Atoian's administration. & Which administrator slots count as being 'under Avetik Atoian's administration': only the primary administrator columns (AdmFName1/AdmLName1), or any of the three administrator slots. & open\\[2pt]
BIRD, \texttt{financial} & What year was account owner number 130 born? & Which ID the number 130 in 'account owner number 130' refers to: the account ID or the client ID & open\\[2pt]
Spider, \texttt{car\_\allowbreak{}1} & What are the makers and models? & Which column supplies 'makers': the integer foreign key model\_list.Maker, or the maker name car\_makers.Maker reached by joining on that key & open\\[2pt]
\midrule
\multicolumn{4}{@{}l}{\textbf{Relationship role} (2 of 6 shown)}\\[1pt]
BIRD, \texttt{toxicology} & Which molecule consisted of Sulphur atom with double bond? Evidence: sulphur refers to element - 's'; double bond refers to bond\_type = ' = '; & Does the sulfur atom itself have to be part of the double bond, or is it enough that the molecule has both a sulfur atom and a double bond somewhere? & open\\[2pt]
Spider, \texttt{network\_\allowbreak{}1} & What is the lowest grade of students who do not have any friends? & Whether a student who appears in Friend only as friend\_id (someone else lists them as a friend) counts as having a friend & open\\[2pt]
\midrule
\multicolumn{4}{@{}l}{\textbf{Strict cutoff} (1 of 1 shown)}\\[1pt]
BIRD, \texttt{thrombosis\_\allowbreak{}prediction} & Provide the ID and age of patient with lactate dehydrogenase (LDH) between 100-300 index above the normal range. Evidence: age refers to SUBTRACT(year(current\_timestamp), year(Birthday)); lactate dehydrogenase (LDH) between 100-300 index above the normal range refers to LDH between 600 and 800; & Whether the LDH range endpoints 600 and 800 are included (strict vs. inclusive bounds) & open\\[2pt]
\midrule
\multicolumn{4}{@{}l}{\textbf{Ties and selection} (3 of 10 shown)}\\[1pt]
BIRD, \texttt{superhero} & Give the hero ID of superhero with the lowest attribute value. Evidence: lowest attribute value refers to MIN(attribute\_value); & How to handle ties at the minimum attribute\_value: return every tied hero, or return just one hero as the singular 'the hero ID' suggests. & open\\[2pt]
BIRD, \texttt{thrombosis\_\allowbreak{}prediction} & For the patient who has the highest Ig A within the normal range, what is his or her diagnosis? Evidence: highest Ig A within the normal range refers to MAX(IGA BETWEEN 80 AND 500); & How to handle several lab records tied at the maximum in-range IgA value (IGA = 500): return one row or every tied patient's diagnosis. & open\\[2pt]
Spider, \texttt{world\_\allowbreak{}1} & Give the name of the nation that uses the greatest amount of languages. & Several countries tie for the most languages (12). The question does not say which of them to return, so the answer depends on how the tie is broken. & settles it\\[2pt]
\midrule
\multicolumn{4}{@{}l}{\textbf{Time scope} (2 of 3 shown)}\\[1pt]
BIRD, \texttt{financial} & For the client who applied the biggest loan, what was his/her first amount of transaction after opened the account? & Whether a transaction on the account-opening date counts as 'after opened the account' (inclusive vs. strictly later boundary) & open\\[2pt]
BIRD, \texttt{thrombosis\_\allowbreak{}prediction} & For all patients with normal uric acid (UA), what is the average UA index based on their latest laboratory examination result? Evidence: uric acid (UA) with normal range refers to UA \textless{} 8.0 and SEX = 'M' or UA \textless{} 6.5 and SEX = 'F'; average UA index refers to AVG(UA) & Which date counts as the 'latest laboratory examination': the single latest date in the whole Laboratory table, or each patient's own latest test date & settles it\\[2pt]
\midrule
\multicolumn{4}{@{}l}{\textbf{Value matching} (3 of 4 shown)}\\[1pt]
BIRD, \texttt{financial} & How much, in total, did client number 617 pay for all of the transactions in 1998? & Which 1998 transactions count as the client having 'paid': every transaction on the client's accounts, or only outgoing ones (debits). & open\\[2pt]
BIRD, \texttt{thrombosis\_\allowbreak{}prediction} & Please list the disease names of the patients that have a proteinuria level higher than normal. Evidence: disease names refers to Diagnosis; proteinuria level higher than normal refers to \texttt{U-PRO} \textgreater{} = 30; & Whether the threshold \texttt{U-PRO} \textgreater{}= 30 compares the TEXT column \texttt{U-PRO} as text (string order) or as numbers & settles it\\[2pt]
Spider, \texttt{car\_\allowbreak{}1} & Which model saves the most gasoline? That is to say, have the maximum miles per gallon. & Whether the TEXT-typed MPG column is ranked by its stored string value or by its numeric value when picking the maximum & settles it\\[2pt]
\end{xltabular}

\begin{xltabular}{\textwidth}{@{}>{\raggedright\arraybackslash}p{1.95cm}>{\hsize=1.1\hsize\raggedright\arraybackslash}X>{\hsize=0.9\hsize\raggedright\arraybackslash}X>{\raggedright\arraybackslash}p{1.45cm}>{\centering\arraybackslash}p{1.0cm}@{}}
\caption{Selected accepted request sets, up to 4 per type.}\label{tab:catsets}\\
\toprule
Set & Open request & Interpretations & Answers & Models\\
\midrule
\endfirsthead
\toprule
Set & Open request & Interpretations & Answers & Models\\
\midrule
\endhead
\bottomrule
\endfoot
\multicolumn{5}{@{}l}{\textbf{Aggregation level} (4 of 61 shown)}\\[1pt]
\texttt{chinook\_\allowbreak{}1}\newline \texttt{1adc54e3cf} & Which media types (by name, counting only tracks that have a media type) have a bytes-per-millisecond rate higher than the simple average of that rate across all media types? For each, show the media type name, its rate, and how much it exceeds that average, unrounded. & \textbf{A:} Each media type's rate is the mean of its tracks' individual Bytes/Milliseconds ratios; list the media types whose rate is above the unweighted mean of all media types' rates, with the rate and how far above it is.\newline \textbf{B:} Each media type's rate is its total Bytes divided by its total Milliseconds; list the media types whose rate is above the unweighted mean of all media types' rates, with the rate and how far above it is. & value 181.7 vs 179.5 & \texttt{-\,B\,B\,N}\\[2pt]
\texttt{formula\_\allowbreak{}1}\newline \texttt{c759056a31} & For each circuit country, what are the average constructor standings points and wins (the season-to-date figures) at races held there? Also, how far is each country's average wins above or below the average of the country figures? & \textbf{A:} A country's averages are the mean of per-race averages, so each race counts equally.\newline \textbf{B:} A country's averages are the plain mean over all of its standings records pooled together. & value 0.039 vs -0.03776 & \texttt{S\,-\,-\,S}\\[2pt]
\texttt{thrombosis\_\allowbreak{}prediction}\newline \texttt{7fd142e2d9} & By patient sex, what are the average aCL IgG and ANA values in the examination data (only examinations linked to a patient, missing values ignored)? Also give each sex's average ANA as a ratio of the mean of the per-sex ANA averages, unrounded. & \textbf{A:} Average aCL IgG and ANA over all matched examination records within each sex. Each examination counts once.\newline \textbf{B:} Average each patient's examinations into per-patient values first. Then average those values within each sex, so each patient counts once. & value 3.281 vs 3.334 & \texttt{S\,S\,S\,S}\\[2pt]
\texttt{alien}\newline \texttt{cde68b2348} & For observatory stations whose average telescope bandwidth usage (\texttt{bandusagepct}) is above average, what is their mean bandwidth usage, and how far above average are they on average? Ignore missing usage values, and give both figures unrounded in a single row. & \textbf{A:} Compare each station's average bandwidth usage with the unweighted mean of the station averages.\newline \textbf{B:} Compare each station's average bandwidth usage with the mean bandwidth usage over all individual telescopes. & value 24.9 vs 25.03 & \texttt{S\,B\,B\,N}\\[2pt]
\midrule
\multicolumn{5}{@{}l}{\textbf{Counting unit} (4 of 47 shown)}\\[1pt]
\texttt{sakila\_\allowbreak{}1}\newline \texttt{fc16192583} & For each actor, show their actor ID, how many films they're credited in, that count as a fraction of the total, and how far that fraction is above or below the average fraction across actors. Return the top 10 by that difference, highest first, unrounded. & \textbf{A:} Each actor's film count is divided by the total number of casting credits in film\_actor, which gives the actor's share of all credits.\newline \textbf{B:} Each actor's film count is divided by the number of distinct films that have a cast in film\_actor, which gives the fraction of those films the actor appears in. & value 0.001591 vs 0.008716 & \texttt{-\,N\,-\,-}\\[2pt]
\texttt{formula\_\allowbreak{}1}\newline \texttt{c1a8ce4d18} & Which circuit cities have an above-average share of qualifying? For each, show the city, its qualifying count, that count as a share of the total across all cities, and how far its share is above the average city share (all unrounded). & \textbf{A:} Each city's count is its number of driver qualifying entries, and its share is its share of all qualifying entries.\newline \textbf{B:} Each city's count is its number of distinct races with qualifying data (qualifying sessions), and its share is its share of all such sessions. & 16 vs 17 rows & \texttt{S\,S\,-\,-}\\[2pt]
\texttt{student\_\allowbreak{}club}\newline \texttt{6b346d7be2} & Taking attendance entries that match an event (others left out), how many attendees does each event \texttt{status} have, and what share of the total across all statuses is that? Show the status, its number of attendees, and the share, without rounding. & \textbf{A:} A status's attendance is the total number of member-event attendance entries for events with that status.\newline \textbf{B:} A status's attendance is the number of different members who attended any event with that status. & value 297 vs 30 & \texttt{S\,B\,S\,S}\\[2pt]
\texttt{insider}\newline \texttt{e6f24d2438} & For each trader type, how many analysed trade records are there in the advanced behaviour data (only where the trade and its trader can be found), and what share of the overall total does each type make up? Keep the share unrounded. & \textbf{A:} For each trader type, count the advanced behaviour analysis records that join to a transaction and a trader. Report that count and its share of the total count over all trader types.\newline \textbf{B:} For each trader type, count the distinct transactions that have at least one advanced behaviour analysis and a matching trader. Report that count and its share of the total count over all trader types. & value 242 vs 241 & \texttt{B\,B\,B\,S}\\[2pt]
\midrule
\multicolumn{5}{@{}l}{\textbf{Missing values} (4 of 29 shown)}\\[1pt]
\texttt{baseball\_\allowbreak{}1}\newline \texttt{f74c2bad9d} & For players who received award votes (matched to \texttt{player} by \texttt{player\_\allowbreak{}id}), what is the average \texttt{points\_\allowbreak{}won} for each death country (ignoring missing points)? Show each death country, its average, and how far it is from the simple average of those per-country averages, without rounding. & \textbf{A:} Average the points won per death country, with a blank death country kept as its own group. Compare each group's average with the plain average of all group averages, the blank group included.\newline \textbf{B:} Average the points won per death country, leaving out players with no recorded death country. Compare each country's average with the plain average of the remaining countries' averages. & 8 vs 7 rows & \texttt{N\,N\,-\,N}\\[2pt]
\texttt{formula\_\allowbreak{}1}\newline \texttt{af91f25689} & Which 10 drivers have the most qualifying entries compared with the average driver? For each, show the driver ID, how many qualifying records they have, and that number divided by the average number of qualifying records per driver (unrounded), highest ratio first. & \textbf{A:} The average number of qualifying records per driver is computed only over drivers who appear in the qualifying table.\newline \textbf{B:} The average number of qualifying records per driver is computed over every driver in the drivers table, with drivers who never qualified counting as zero. & value 4.112 vs 22.63 & \texttt{S\,B\,B\,B}\\[2pt]
\texttt{student\_\allowbreak{}club}\newline \texttt{55be105b98} & Group members by their major (\texttt{link\_\allowbreak{}to\_\allowbreak{}major}) and show each major id, how many members it has, and that count's share of the total. List the top 10 by share, highest first, with no rounding. & \textbf{A:} Members without a major form their own group, and they are included in the total used as the denominator.\newline \textbf{B:} Members without a major are left out of both the groups and the total used as the denominator. & 10 rows; 1 differ & \texttt{B\,B\,B\,N}\\[2pt]
\texttt{crypto}\newline \texttt{b61f4a52bd} & For each order in \texttt{orderexecutions.\allowbreak{}ordersmark} (executions with no order reference grouped together), what's the average size (\texttt{fillcount}) of the fills it got? Keep the orders whose average beats the average across all orders, and note by how much. Then, among those, keep the ones that also beat that group's own average, and note that margin too. Give me one summary row with three values: the average per-order fill size of the final orders, their average margin over the first benchmark, and their average margin over the second. Leave everything unrounded. & \textbf{A:} Zero-quantity executions count as fills of 0 in each order's average fill quantity.\newline \textbf{B:} Zero-quantity executions are not fills and are left out of each order's average fill quantity. & value 4.024 vs 4.021 & \texttt{S\,S\,S\,S}\\[2pt]
\midrule
\multicolumn{5}{@{}l}{\textbf{Negation} (1 of 1 shown)}\\[1pt]
\texttt{customers\_\allowbreak{}and\_\allowbreak{}invoices}\newline \texttt{2e529bb193}\newline (filter) & Ignoring transactions with no invoice number, which invoice numbers have transactions without the type 'Payment' (exact text)? Show each invoice number with the total number of transactions recorded for it. Sort by that count from highest to lowest and give the top 10. & \textbf{A:} Invoices with no Payment transaction at all. For each one, count its transactions and show the top 10 by count.\newline \textbf{B:} Invoices with at least one transaction that is not a Payment. For each one, count all its transactions (Payments included) and show the top 10 by count. & 4 vs 7 rows & \texttt{S\,S\,N\,S}\\[2pt]
\midrule
\multicolumn{5}{@{}l}{\textbf{Strict cutoff} (4 of 42 shown)}\\[1pt]
\texttt{chinook\_\allowbreak{}1}\newline \texttt{06934ae88f}\newline (filter) & Using 3.98 as the cutoff on the invoice total (\texttt{Invoice.\allowbreak{}Total}) to keep only the higher-value invoices, what is their average total, as a single unrounded value? Ignore missing totals, and give a missing result if no invoices qualify. & \textbf{A:} Invoices whose total equals 3.98 are included in the average.\newline \textbf{B:} Only invoices whose total is strictly greater than 3.98 are included in the average. & 1 rows; 1 differ & \texttt{B\,B\,B\,S}\\[2pt]
\texttt{financial}\newline \texttt{8a55588ad2}\newline (filter) & Applying a cutoff of 2802.0 to the debited amount (\texttt{order.\allowbreak{}amount}), what is the average debited amount of the large standing orders, unrounded? & \textbf{A:} Orders of exactly 2802.0 are included.\newline \textbf{B:} Only orders above 2802.0 are included. & 1 rows; 1 differ & \texttt{B\,B\,B\,B}\\[2pt]
\texttt{formula\_\allowbreak{}1}\newline \texttt{57277c47b4}\newline (filter) & Among driver standings entries where the driver has upwards of 1 point so far that season, what is the average season-to-date win count? Ignore missing values and don't round. & \textbf{A:} Average the season-to-date win counts over standings entries whose season-to-date points are 1 or more.\newline \textbf{B:} Average the season-to-date win counts over standings entries whose season-to-date points are more than 1. & 1 rows; 1 differ & \texttt{B\,B\,B\,S}\\[2pt]
\texttt{insider}\newline \texttt{8ed79619ea}\newline (filter) & Using 0.149 as the cutoff, how many advanced behaviour records have high peer correlation (\texttt{peercorr})? Give a single count. & \textbf{A:} Peer correlation values equal to 0.149 are counted (at least 0.149).\newline \textbf{B:} Only peer correlation values strictly greater than 0.149 are counted (more than 0.149). & 1 rows; 1 differ & \texttt{B\,B\,B\,N}\\[2pt]
\midrule
\multicolumn{5}{@{}l}{\textbf{Ties and selection} (4 of 24 shown)}\\[1pt]
\texttt{sakila\_\allowbreak{}1}\newline \texttt{b77a784ee8} & What are the top 10 films by number of copies in our inventory? For each, show the \texttt{film\_\allowbreak{}id}, how many copies we have, and what share of all inventory copies that is (copies divided by the total number of copies, unrounded), with the highest share first. & \textbf{A:} List the top 10 films by share of inventory copies, returning exactly 10 rows and choosing arbitrarily among films tied at the cut.\newline \textbf{B:} List the top films by share of inventory copies, keeping every film whose copy count reaches the 10th-highest count, so ties at the cut are all included. & 10 vs 72 rows & \texttt{-\,N\,-\,-}\\[2pt]
\texttt{financial}\newline \texttt{53afca1d49} & What are the top 10 branch districts (\texttt{account.\allowbreak{}district\_\allowbreak{}id}) by share of all accounts? For each, show the district id, its number of accounts and its share of all accounts, without rounding, listed from the highest share down. & \textbf{A:} Rank districts by their share of accounts and return exactly the first 10 rows, dropping tied districts arbitrarily at the cut.\newline \textbf{B:} Rank districts by their share of accounts and return every district whose share reaches the 10th-highest share, including all ties at the cut. & 10 vs 12 rows & \texttt{S\,N\,S\,S}\\[2pt]
\texttt{student\_\allowbreak{}club}\newline \texttt{5c048dd871} & Which are the top 10 members by share of total event attendance? Match attendance to members by member id, count attendance per member id and position, and show the member id, position, attendance count and share of all attendance, from highest share to lowest. & \textbf{A:} Rank member/position groups by their share of all attendance and return exactly 10 rows, cutting arbitrarily through any tie at 10th place.\newline \textbf{B:} Rank member/position groups by their share of all attendance and return the top 10 plus every row tied with the 10th row's share. & 10 vs 12 rows & \texttt{S\,N\,S\,S}\\[2pt]
\texttt{alien}\newline \texttt{ee6b268047} & What are the top 10 telescopes by share of detected signals? For each telescope reference (\texttt{telescref}), show how many signals it detected and what fraction of all signals that is (unrounded), from the largest share down. & \textbf{A:} Rank telescopes by their share of all signals and return exactly 10 rows, choosing arbitrarily among telescopes tied at the 10th place.\newline \textbf{B:} Rank telescopes by their share of all signals and return every telescope whose share is at least the 10th-highest share, including all ties at the cut. & 10 vs 51 rows & \texttt{-\,N\,S\,-}\\[2pt]
\end{xltabular}
}
\twocolumn

\end{document}